\documentclass[journal]{IEEEtran}
\pdfoutput=1
\usepackage[T1]{fontenc}
\usepackage{ifpdf}
\usepackage{cite}
\usepackage[pdftex]{graphicx}
\usepackage[cmex10]{amsmath}
\usepackage{bm}
\usepackage{algorithmic}
\usepackage{array}
\usepackage[cmintegrals]{newtxmath}
\usepackage{algorithmic}
\usepackage{fixltx2e}
\usepackage{dblfloatfix}
\usepackage{multicol}
\usepackage{float}
\usepackage{color}
\usepackage{layouts}
\usepackage{algorithm,algorithmic}
\usepackage{tikz}

\usepackage{graphicx}
\usepackage{epstopdf}
	\DeclareGraphicsExtensions{.eps}
\usepackage{multicol}
\usepackage{caption}
\usepackage{subcaption}
\usepackage{mwe}
\usepackage{enumitem}
\usepackage{standalone}

\newtheorem{theorem}{Theorem}

\newtheorem{lemma}{Lemma}
\newtheorem{proposition}{Proposition}

\DeclareMathOperator*{\argmax}{arg\,max}

\newcommand{\bS}{\mathbb{S}}

\usetikzlibrary{shapes.geometric}
\usetikzlibrary{calc}
\usetikzlibrary{positioning}

\begin{document}


\title{Learning to Bound the Multi-class Bayes Error}

\author{Salimeh~Yasaei Sekeh$^{\ast\dagger}$, Member, IEEE,
        Brandon Oselio$^{\dagger}$, Member, IEEE,\\
        ~Alfred~O.~Hero$^{\dagger}$, Fellow, IEEE~
\\
$^{\dagger}$Department of EECS, University of Michigan, Ann Arbor, MI, USA\\
$^{\ast}$School of Computing and Information Science, University of Maine, Orono, ME, USA.
\thanks{The work presented in this paper was partially supported by ARO grant W911NF-15-1-0479 and DOE grant DE-NA0002534. The authors also acknowledge support from AFRL under grant USAF FA8650-15-D-1845. Some parts of this paper have been presented at the Annual Allerton Conference on Communication, Control, and Computing, 2018.}

}

\maketitle

\begin{abstract}
In the context of supervised learning, meta learning uses features, metadata and other information to learn about the difficulty, behavior, or composition of the problem. Using this knowledge can be useful to contextualize classifier results or allow for targeted decisions about future data sampling. In this paper, we are specifically interested in learning the Bayes error rate (BER) based on a labeled data sample. Providing a tight bound on the BER that is also feasible to estimate has been a challenge. Previous work~\cite{BerishaICassp2016} has shown that a pairwise bound based on the sum of Henze-Penrose (HP) divergence over label pairs can be directly estimated using a sum of Friedman-Rafsky (FR) multivariate run test statistics. However, in situations in which the dataset and number of classes are large, this bound is computationally infeasible to calculate and may not be tight. Other multi-class bounds also suffer from computationally complex estimation procedures. In this paper, we present a generalized HP divergence measure that allows us to estimate the Bayes error rate with log-linear computation. We prove that the proposed bound is tighter than both the pairwise method and a bound proposed by Lin \cite{Lin1991}. We also empirically show that these bounds are close to the BER. We illustrate the proposed method on the MNIST dataset, and show its utility for the evaluation of feature reduction strategies. We further demonstrate an approach for evaluation of deep learning architectures using the proposed bounds.

\end{abstract}

\begin{IEEEkeywords}
meta learning, multi-class classification, Henze-Penrose divergence, Bayes error rate, Friedman-Rafsky statistic, minimal spanning tree.
\end{IEEEkeywords}

\IEEEpeerreviewmaketitle
\def\BX{\mathbf{X}}
\def\BS{\mathbf{S}}
\def\BK{\mathbf{K}}
\def\tL{\mathbf{L}}
\def\BB{\mathbf{B}}
\def\vphi{{\varphi}}
\def\rw{{\rm w}}
\def\bZ{\mathbf Z}
\def\wtf{{\widetilde f}} \def\wtg{{\widetilde g}} \def\wtG{{\widetilde G}}
\def\vphi{\varphi}
\def\rT{{\rm T}}
\def\tA{{\tt A}} \def\tB{{\tt B}} \def\tC{{\tt C}} \def\tI{{\tt I}} \def\tJ{{\tt J}} \def\tK{{\tt K}}
\def\tL{{\tt L}} \def\tP{{\tt P}} \def\tQ{{\tt Q}} \def\tS{{\tt S}}
\def\beac{\begin{array}{c}} \def\beal{\begin{array}{l}} \def\beacl{\begin{array}{cl}} \def\ena{\end{array}}
\def\bbV{\mathbb{V}}
\def\bbS{\mathbb{S}}
\def\diy{\displaystyle}
\def\bx{\mathbf{x}}
\def\rd{\rm {d}}
\def\ep{\epsilon}
\def\bbR{\mathcal{R}}
\def\bbE{\mathcal{E}}

\ifCLASSOPTIONcaptionsoff
  \newpage
\fi

\section{Introduction}

\IEEEPARstart{M}{}eta-learning is a method for learning the intrinsic quality of data directly from a sample of the data, metadata, or other information \cite{ChanStolfo1993}, \cite{Prodromidisetal2000}. The purpose of meta-learning is to collect knowledge that might be helpful at other levels of processing and decision-making. Examples where meta-learning is applied include sequential design of experiments \cite{CarneiroandVasconcelos2005}, reinforcement learning \cite{Guptaetal2018}, and sensor management \cite{Xieetal2017} in the fields of statistics, machine learning, and systems engineering, respectively. In supervised learning, and particularly for multi-class classification, one form of meta-learning is to learn bounds or estimates on the Bayes error rate (BER). The BER is the minimal achievable error probability of any classifier for the particular learning problem, and knowledge of it can be used at other stages of meta-learning, such as in the selection of the classifiers, model selection, and feature reduction. Hence, finding computable bounds and approximations to the BER is of interest, and is the problem we consider in this paper. 

Consider the problem where a feature vector $\BX$ is labeled over $m$ classes $C_1,\ldots,C_m$.  Available are i.i.d. pairs $\{(\mathbf{x}_i, y_i)\}_{i=1}^n$, called training data, where $\mathbf{x}_i$ is a realization of the random vector (feature) $\mathbf{X} \in \mathbb{R}^d$ and $y_i$ is a realization of the random variable (label) $Y\in\{1,2,\dots,m\}$. Assume the prior label probabilities $p_k=P(Y=k)$, with $\diy\sum\limits_{k=1}^m p_k=1$ and the conditional feature densities $f_k(\bx)=f(\bx|Y=k)$, for $k=1,\dots,m$. Then the Bayes error rate is given by
\begin{equation}\label{BER.m-classes}
\epsilon^m=1-\diy\int \max\{p_1f_1(\bx),p_2f_2(\bx),\dots,p_m f_m(\bx)\}\;\rd\bx.
\end{equation}
This represents the error achieved by the Bayes classifier, $g_{Bayes}$ that minimizes the average $0-1$ loss. The Bayes classifier assigns an estimated class label $\hat{y}$ to an observation $\bx$ according to the maximum a posterior (MAP) rule
\begin{equation*}
\hat{y} = g_{Bayes}(\bx) = \argmax_{k \in \{1, 2, \ldots, m\}} P(Y=k | \mathbf{X} = \bx).
\end{equation*}

Many different upper and lower bounds on the BER (\ref{BER.m-classes}) exist for the case of $m=2$ classes, and many of these are related to the family of $f$-divergences. A bound based on Chernoff $\alpha$-divergence has been proposed in \cite{Chernoff1952}, but in general it is not very tight in the finite sample regime. In \cite{Berishaetal2016}, a tighter bound for the 2-class BER using Henze-Penrose \cite{HP} divergence was proposed. The HP bound has the advantage that it can be directly estimated from the training data using a minimal spanning tree. The same framework can be extended in a pairwise fashion to the $m$-class multi-class classification problem. However, when $m$ is relatively large, the derived pairwise bounds are loose and often times trivial~\cite{YOH.Allerton2018}. The method proposed in this paper alleviates this problem by introducing new bounds based on a generalized Henze-Penrose measure adapted to $m$-class problem, and whose tightness does not diminish as $m$ increases. Additionally, the new bounds improve upon other bounds that were designed specifically for the multi-class problem, such as the generalized Jensen-Shannon (JS) divergence bound~\cite{Lin1991}.

Most approaches to estimation of bounds on Bayes error use plug-in estimators. These approaches require estimation of the multivariate feature densities followed by evaluation of the BER bounds using these estimated densities in place of the true densities. Recently, approaches to estimating BER bounds using direct estimators have been proposed. For example, graph-based BER bound estimation approaches bypass density estimation entirely, producing an estimator of the information divergence using geometric functions of the data. These procedures scale linearly in dimension, resulting in faster computation than plug-in methods for high dimensional features. In the original 2-class setting, as shown in \cite{Berishaetal2016}, bounds based on Henze-Penrose divergence can be estimated directly from data by employing the Friedman-Rafsky (FR) run test statistic~\cite{FR,HP}, which computes a minimal spanning tree (MST) over the data, and counts the number of edges that connect dichotomous data points. A brute force extension of the FR approach to the $m$-class classification problem would require an MST computation for each pair of classes, or $O(m^2)$ MSTs, which significantly reduces its computational tractability for large $m$. The extension proposed in this paper also uses a graph-based estimation procedure, but only requires a single MST calculation on the entire dataset. Thus, the proposed approach is more computationally efficient when $m$ and $n$ are large.

\subsection{Related Work}

Broadly defined, meta-learning is a set of methods of learning from knowledge that can be used to improve performance or understanding of the problem. Estimating the Bayes multi-class classification error is a meta-learning problem. The principles behind the frameworks proposed in \cite{LissackandFu1976} and \cite{GarberDjouadi1988} have been utilized to estimate the multi-class BER by bounding the BER by a sum of pairwise BERs for each pair of class labels. There exist many useful bounds on the pairwise BER that are based on information divergence measures, i.e., measures of dissimilarity between two distributions. Several bounds for the pairwise BER have been proposed, including: Chernoff bound \cite{Chernoff1952};  Bhattacharyya bound \cite{Kailath1967}; and HP-divergence \cite{Berishaetal2016}. The Henze-Penrose divergence yields tighter bounds on the BER than those based on the Bhattacharya distance for equal label priors. For the multi-class BER, the sum of pairwise bounds given in \cite{BerishaICassp2016} was proposed. 

Another approach to bounding the BER of multi-class classifiers uses the Jensen-Shannon (JS) divergence. 
The JS-divergence assigns a different weight to each conditional probability distribution and this inspired Lin to propose a bound for the binary BER where the weights depend on the priors.
The generalized multi-class Jensen-Shannon divergence is related to the Jensen difference proposed by Rao \cite{Rao1985,Rao1982}. In \cite{Lin1991}, the author proposed a generalized JS-divergence that was used to derive a bound on the Bayes error rate for multi-class classification.

In the nonparametric setting, the most popular approach for estimating bounds has been plug-in estimators, which require estimation of probability densities that are subsequently ``plugged into'' the expression for the divergence function, \cite{Moon2014ISIT, Moon2016ISIT, MSGH}. These approaches are generally multi-step procedures and computationally expensive, especially in high dimensions \cite{MoonHero2014}. In \cite{Nguyenetal2010}, Nguyen et al. proposed a divergence estimation method based on estimating the likelihood ratio of two densities that achieves the parametric mean squared error (MSE) convergence rate when the densities are sufficiently smooth. 

Direct estimation approaches bypass density estimation, producing an estimator
of the information divergence using geometric functions of the data. As
previously mentioned, the MST-based Friedman-Rafsky two sample test
statistic~\cite{FR,HP} is an asymptotically consistent estimator of the
HP-divergence, which can then be used to estimate upper and lower bounds for the
2-class classification problem~\cite{Berishaetal2016}. There are other
graph-based estimators that have been proposed in the literature. The author in \cite{Henze1988} proposed a graph-based estimator for HP-divergence that employs the K-nearest neighbor (K-NN) graph instead of the MST. The authors of \cite{MNYSH2017} developed an approach for estimating general $f$-divergences called the Nearest Neighbor Ratio (NNR), also using K-NN graphs. In \cite{noshad2018scalable} the authors developed a general divergence estimator based on Locality Sensitive Hashing (LSH). 
In \cite{YOH}, the authors showed that a cross match statistic based on optimal weighted matching can also be used to directly estimate the HP divergence. None of these papers on geometric methods proposed extensions to multi-class classification, which is the main contribution of this paper. 
\subsection{Contribution}
We introduce a computationally scalable and statistically consistent method for learning to bound the multi-class BER. First, we propose a novel measure, the generalized Henze-Penrose (GHP) integral, for bounding multi-class BER. We show how this generalized integral can be used to calculate bounds on the Bayes error rate, and prove that they are tighter than both the pairwise and JS multi-class BER bounds. Further, we empirically show that the bounds' performance is consistent and is robust to sample size and the number of classes.

 Our second contribution is a scalable method for estimating the GHP integral, and subsequent estimation of the BER bounds. The proposed algorithm uses a single global minimal spanning tree (MST) constructed over the entire dataset. We show that this is more computationally efficient than the pairwise approach, which must compute $O(m^2)$ pairwise MSTs. 
\subsection{Organization of the paper}
The paper is organized as follows. In Section~\ref{sec.HP-div} we briefly review the HP divergence and propose the generalized HP-integral (GHP) measure. The motivation and theory for the various bounds such as the pairwise HP divergence and generalized JS divergence for the multi-class Bayes error is reviewed in Section~\ref{Sec.BERbounds}, and a new bound based on our GHP measure is given. We numerically illustrate the theory in Section~\ref{sec.simulation}. In Section~\ref{Real-dataset} we apply the proposed method to a real dataset,  the MNIST image dataset. Finally, Section~\ref{sec:conclusion} concludes the paper. The main proofs of the theorems in the paper are given in the longer version of the paper on arXiv. 
\section{The Divergence Measure and Generalizations}

In this section we recall the Henze-Penrose (HP) divergence between pairs of densities and define a generalization for multiple densities ($\ge 2$) that will be the basis for learning to bound multi-class BER, called "Generalized HP-integral". 
\subsection{Henze-Penrose Divergence}\label{sec.HP-div}
\def\BX{\mathbf{X}}
For parameters $p\in(0,1)$ and $q=1-p$ consider two density functions $f$ and $g$ with common domain $\mathbb{R}^d$. The Henze-Penrose divergence $D(f,g)$ is given by 
\begin{equation}\label{HP-divergence}
    D(f,g)=\diy\frac{1}{4pq}\left[\int \diy\frac{\big(p f(\bx)-qg(\bx)\big)^2}{pf(\bx)+qg(\bx)}\;{\rm d}\bx-(p-q)^2\right].
\end{equation}
The HP-divergence (\ref{HP-divergence}), first introduced in \cite{BerishaHero2015}, has the following properties: (1) $0\leq D \leq 1$, (2) $D=0$ iff $f(\bx)=g(\bx)$. Note that the HP-divergence belongs to the $f$-divergence family \cite{Csisz1967, AliS1966, Morimoto1963}. 

In the multi-class classification setting, as defined in the introduction, consider a sample of i.i.d. pairs  $(\mathbf{x}_i, y_i)$, $i=1, \ldots, n$, where $y_i \in \{1, \ldots, m\}$ are class labels with prior probabilities $\{p_k\}_{k=1}^m$ and, given $y_i=k$, $\mathbf{x}_i$ has conditional density $f_k$.
Define $\widetilde{p}_{ij}=p_i/(p_i+p_j)$. Note that $\widetilde{p}_{ij}\neq \widetilde{p}_{ji}$ and $\widetilde{p}_{ij}+\widetilde{p}_{ji}=1$. Let $\bbS^{(i)}$ be the support set of the conditional distribution $f_i$.  
The Henze-Penrose (HP) divergence measure between distributions $f_{i}$ and $f_{j}$ with union domain $\bbS^{(ij)}=\bbS^{(i)}\cup \bbS^{(j)}$ is defined as follows (see \cite{HP,BerishaHero2015,Berishaetal2016}):
\begin{align}\label{EQ:DP} 
& D(f_{i},f_{j})\nonumber\\
&\quad=\diy\frac{1}{4\widetilde{p}_{ij}\widetilde{p}_{ji}}\Big[\int_{\bbS^{(ij)}} \frac{\big(\widetilde{p}_{ij} f_{i}(\bx)-\widetilde{p}_{ji} f_{j}(\bx)\big)^2}{\widetilde{p}_{ij} f_{i}(\bx)+\widetilde{p}_{ji} f_{j}(\bx)}\;\rd \bx -({\it \widetilde{p}_{ij}-\widetilde{p}_{ji})}^2\Big].
\end{align}
An alternative form for $D(f_{i}, f_{j})$ is given in terms of the HP-integral:
\begin{equation}\label{Def:HP}
{\rm HP}_{ij}:={\rm HP}\big(f_i,f_j\big)=\int_{\bbS^{(ij)}}\diy\frac{f_{i}(\bx)f_{j}(\bx)}{p_i f_{i}(\bx)+p_j f_{j}(\bx)}\rd\bx,\end{equation}
yielding the equivalent form to (\ref{EQ:DP})
$$D(f_{i},f_{j})=1-(p_i+p_j){\rm HP}_{ij}.$$
In \cite{Berishaetal2016} it was shown that $0\leq D(f_{i},f_{j}) \leq 1$, and that the HP-integral is upper bounded by ${\rm HP}\big(f_i,f_j\big)\leq (p_i+p_j)^{-1}$.

\subsection{Generalized HP-Integral}
Define the union of all support sets as $\bbS=\bigcup\limits_{k=1}^m \bbS^{(k)}$ and the difference between the $m$-fold support set and the $2$-fold support set ${\bbS}^{(ij)}$ as $\overline{\bbS}^{(ij)}=\bbS\big/\bbS^{(ij)}$.
We denote $f^{(m)}(\bx)$ the marginal distribution of $\BX$,
$$f^{(m)}(\bx):=\diy\sum\limits_{k=1}^m p_k f_{k}(\bx)=\diy\sum\limits_{k=1}^m p_k f(\bx|y=k). $$
Define the generalized HP-integral (GHP-integral) by 
\begin{equation}\label{Def:GHP}
{\rm GHP}^m_{ij}:={\rm GHP}^m\big(f_{i},f_{j}\big)=\diy\int_{\bbS}\diy f_{i}(\bx)f_{j}(\bx)\big/f^{(m)}(\bx)\;\rd\bx.\end{equation}
Note that GHP-integral is not a linear function of a divergence function. Now the question is when a random vector with $m$ class label is considered what is relation between ${\rm HP}_{ij}$ and ${\rm GHP}^m_{ij}$. This explicitly is discussed below.  
\subsection{A Relation Between GHP-Integral and HP-Integral}
Consider conditional probability densities $f_1,f_2,\dots,f_m$ with priors $p_1,p_2,\dots,p_m$ such that $p_1+p_2+\dots+p_m=1$. 
The HP-integral and the GHP-integral are related as follows: 
\begin{itemize}
\item[(a)] If $\left(\bbS^{(i)}\cup\bbS^{(j)}\right)\cap \bigcup\limits_{k\neq i,j}\bbS^{(k)}=\emptyset$, then 
\begin{equation} \label{eq02:thm1} {\rm HP}(f_{i},f_{j})={\rm GHP}^m\big(f_{i},f_{j}\big), \end{equation}
\item[(b)] If $\left(\bbS^{(i)}\cup\bbS^{(j)}\right)\cap \bigcup\limits_{k\neq i,j}\bbS^{(k)}\neq\emptyset$, then there exists a constant C depending only on priors $p_1,p_2,\dots,p_m$ such that 
\begin{equation}\label{eq03:thm1} \diy {\rm HP}(f_{i},f_{j})\leq {\rm GHP}^m\big(f_{i},f_{j}\big)
+\diy C\Big(1-D\Big(\widetilde{p}_{ij}f_i+\widetilde{p}_{ji}f_j,\diy\sum\limits_{k\neq i,j}\widetilde{p}_k^{ij}f_k\Big)\Big), \end{equation}
where $\widetilde{p}_{ij}=p_i/(p_i+p_j)$ and $\widetilde{p}_k^{ij}=p_k\big/\diy\sum\limits_{r\neq i,j}p_r.$
\end{itemize}

An analytical discussion on relations (\ref{eq02:thm1}) and (\ref{eq03:thm1}) is given in Appendix-A. 
Intuitively we imply that the HP divergence in (\ref{eq03:thm1}) increases when the support set of samples with labels $i$ and $j$ are nearly disjoint from the support sets of the other labels $k\neq i,j$ $k=1,\dots,m$. In this case the HP-integral becomes closer to the GHP-integral. Specially, (\ref{eq03:thm1}) approaches (\ref{eq02:thm1}) as the intersection between support sets $\bbS^{(ij)}$ and $\bigcup\limits_{k\neq i,j}\bbS^{(k)}$ decreases, i.e. the conditional distributions become less overlapping. This in fact validates the relation (7) intuitively.

\section{BOUNDS ON THE BAYES ERROR RATE}\label{Sec.BERbounds}
Before introducing the new bound on multi-class BER, we first review the pairwise bounds on the multi-class Bayes error rate given by Berisha et al. \cite{BerishaICassp2016} and by Lin \cite{Lin1991}. 
\subsection{Pairwise HP Bound}
For the case of $m$ classes the authors in \cite{BerishaICassp2016} have shown that the multi-class BER $\epsilon^m$ in (\ref{BER.m-classes}) can be bounded by
\begin{equation}\label{multiclass.HP-bound}\begin{array}{l}
\diy\frac{2}{m}\diy\sum\limits_{i=1}^{m-1}\sum\limits_{j=i+1}^m (p_i+p_j)\epsilon_{ij}\leq \epsilon^m\leq \diy\sum\limits_{i=1}^{m-1}\sum\limits_{j=i+1}^m (p_i+p_j)\epsilon_{ij},
\end{array}\end{equation}
where $\epsilon_{ij}$ represents the pairwise Bayes risk of the two class sub-problem of classifying between classes $i$ and $j$:
\begin{equation}
\epsilon_{ij}=\diy\int \min \big\{\widetilde{p}_{ij}f_{i}(\bx),\widetilde{p}_{ji}f_{j}(\bx)\big\}\;\rd\bx.
\end{equation}
In \cite{Berishaetal2016}, it has been shown that
\begin{equation}\label{Berisha.bound}
\diy\frac{1}{2}-\frac{1}{2} \sqrt{u_{\widetilde{p}_{ij}}(f_{i},f_{j}})\leq \epsilon_{ij}\leq \diy\frac{1}{2}-\frac{1}{2}u_{\widetilde{p}_{ij}}(f_{i},f_{j}), 
\end{equation}
\begin{equation}\label{expression:U}
\hbox{where}\;\;u_{\widetilde{p}_{ij}}(f_{i},f_{j})=4\widetilde{p}_{ij}\widetilde{p}_{ji}\; D(f_{i},f_{j})+(\widetilde{p}_{ij}-\widetilde{p}_{ji})^2,
\end{equation}
and $D(f_i,f_j)$ is defined in (\ref{EQ:DP}). Using both (\ref{multiclass.HP-bound}) and (\ref{Berisha.bound}), we obtain bounds for the multi-class Bayes error rate. While these bounds have been successfully applied \cite{FR}, \cite{Berishaetal2016}, it has the disadvantage of high computational complexity due to the presence of ${m \choose 2}$ summands in (\ref{multiclass.HP-bound}).
\def\bbE{\mathbb{E}}
\subsection{JS Bound}
The generalized Jensen-Shannon divergence is defined as 
$${JS}_p(f_1,f_2,\dots,f_m)=\bar{H}\left(\diy\sum\limits_{k=1}^m p_kf_k\right)-\diy\sum\limits_{k=1}^m p_i\bar{H}(f_i), $$
where $\bar{H}$ is the Shannon entropy function.
In \cite{Lin1991} this divergence measure was used to obtain a bound on the multi-class BER. The Bayes error rate $\epsilon^m$ is upper bounded by
\begin{equation}\label{Upper-Lin}
\epsilon^m\leq\diy\frac{1}{2}\big(H(p)-{JS}_p(f_1,f_2,\dots,f_m)\big):=\epsilon^m_{\rm JS} \;\;\hbox{(say)},
\end{equation}
and is lower bounded by 
\begin{equation}\label{JS:lowerbound}
\epsilon^m\geq \diy\frac{1}{4(m-1)}\big(H(p)-{JS}_p(f_1,f_2,\dots,f_m)\big)^2.
\end{equation}
Here $H(p)$
is Shannon entropy and $JS$ is generalized Jensen-Shannon divergence. The bounds in (\ref{Upper-Lin}) and (\ref{JS:lowerbound}) can be approximated by plug-in estimation or by direct methods, such as the NNR method \cite{NMYH} or other graph methods \cite{Heroetal2002}. We will show that the JS bound suffers from lack of tightness.


\subsection{Proposed Multi-class Bayes Error Probability Bound}
 
To simplify notation, denote
$$
\delta_{ij}:=\diy\int \diy\frac{p_i p_j f_i(\bx)f_j(\bx)}{p_if_i(\bx)+p_jf_j(\bx)}\rd\bx,\;\;\;{\it\delta^m_{ij}:= \diy\int \diy\frac{p_i p_j f_i(\bx)f_j(\bx)}{f^{(m)}(\bx)}\rd\bx}, 
$$

and note that 
$ \delta_{ij} = \frac{(p_i+p_j)}{4}\big(1-u_{\widetilde{p}_{ij}}(f_{i},f_{j})\big)$, where $u_{\widetilde{p}_{ij}}$ is defined in (\ref{expression:U}) and $\delta_{ij} \geq \delta_{ij}^{m}$.

\begin{theorem}\label{thm.2} 
{\it For given priors $p_1,p_2,\dots,p_m$ and conditional distributions $f_1,f_2,\dots,f_m$, 
the multi-class BER $\ep^m$ satisfies 
\begin{equation}\label{proposed.upbound}
\ep^m\leq 2\sum\limits_{i=1}^{m-1}\sum\limits_{j=i+1}^m \delta^m_{ij}.
\end{equation}
And is lower bounded by $\delta^m_{ij}$ as
\begin{equation}\label{lowerbound.2}
\ep^m\geq \diy\frac{m-1}{m}\left[ 1 - \left(1 - 2\frac{m}{m-1} \sum_{i=1}^{m-1} \sum_{j=i+1}^{m} \delta^m_{ij}\right)^{1/2}\right].
\end{equation}
}
\end{theorem}
In the following theorem we show that the proposed upper and lower bounds are tighter than the JS upper (\ref{Upper-Lin}) and lower (\ref{JS:lowerbound}) bounds.

\begin{theorem}\label{GHPandJS.upbound}
{\it For given priors $p_1,p_2,\dots,p_m$ and conditional distributions $f_1,f_2,\dots,f_m$, for $m\geq 3$
\begin{equation}\label{Ineq:tighterJS}
\ep^m\leq 2\sum\limits_{i=1}^{m-1}\sum\limits_{j=i+1}^m \delta^m_{ij}\leq
\epsilon^m_{\rm JS},
\end{equation}
\begin{equation}\label{lowerbound.1.tight}\begin{array}{l}
\ep^m\geq \diy\frac{m-1}{m}\left[ 1 - \left(1 - 2\frac{m}{m-1} \sum_{i=1}^{m-1} \sum_{j=i+1}^{m} \delta^m_{ij}\right)^{1/2}\right]\geq \diy \frac{(\epsilon^m_{\rm JS})^2}{(m-1)}.
\end{array}\end{equation}
}
\end{theorem}

Theorem~\ref{thm.lower.tight.2} shows that proposed upper and lower bounds are tighter than bounds in (\ref{multiclass.HP-bound}), i.e., the pairwise (PW) bounds. 
\begin{theorem}\label{thm.lower.tight.2}
{\it For given priors $p_1,p_2,\dots,p_m$ and conditional distributions $f_1,f_2,\dots,f_m$, the multi-class classification BER $\ep^m$ is upper bounded 
\begin{equation}\label{upper.toghterPW}
\ep^m\leq 2\sum\limits_{i=1}^{m-1}\sum\limits_{j=i+1}^m \delta^m_{ij}\leq 2\diy\sum\limits_{i=1}^{m-1}\sum\limits_{j=i+1}^m \delta_{ij}. 
\end{equation}
and is lower bounded by $\delta^m_{ij}$ as
\begin{equation}\label{lowerbound.2.tight}\begin{array}{l}
\ep^m\geq \diy\frac{m-1}{m}\left[ 1 - \left(1 - 2\frac{m}{m-1} \sum_{i=1}^{m-1} \sum_{j=i+1}^{m} \delta^m_{ij}\right)^{1/2}\right]\\
\\
\qquad \geq \diy\frac{2}{m}\diy\sum\limits_{i=1}^{m-1}\sum\limits_{j=i+1}^m (p_i+p_j)\left[\diy\frac{1}{2}-\frac{1}{2} \sqrt{u_{\widetilde{p}_{ij}}(f_{i},f_{j}})\right].
\end{array}\end{equation}
where $u_{\widetilde{p}_{ij}}$ is given in (\ref{expression:U}). 
}
\end{theorem}

The proofs of Theorems \ref{thm.2}, \ref{GHPandJS.upbound}, and
\ref{thm.lower.tight.2} are given in Appendix B, C, and D respectively.

In addition to the above Theorem, we are able to more precisely characterize the
relationship between the pairwise and GHP upper bounds:

\begin{proposition}
  \label{prop:1}
  The equality on the right hand side of (\ref{upper.toghterPW}) holds
iff a) $m=2$, or b) $\forall i \neq j, (\bS^{(i)} \cup
\bS^{(j)}) \cap \cup_{k \neq i, j} \bS^{(k)}=\emptyset$.
\end{proposition}


\begin{proposition}
  \label{prop:2}
   The equality on the right hand side of (\ref{lowerbound.2.tight}) holds
iff $m=2$.
\end{proposition}

The proofs of Propositions \ref{prop:1} and \ref{prop:2} are given in Appendix E and F.
\section{Learning the bounds from data}\label{sec.estimate.delta}

Here we review the pairwise Friedman-Rafsky (FR) statistic and introduce a generalized FR statistic. Given a labeled sample $\BX=\{(\mathbf{x}_i, y_i)\}_{i=1}^n$ define the subset of samples having label $k$ as: $\mathbf{X}^{(k)}=\{(\mathbf{x}_i, y_i)\}_{i=1,y_i=k}^n$, $k=1, \ldots, m$.  The cardinality of the subset $\mathbf{X}^{(k)}$ is $n_k=\sum_{i=1}^n I(y_i=k)$ where $I(B)$ denotes the indicator function of event $B$.  
We denote the pairwise FR statistic by $\mathfrak{R}_{n_i,n_j}$ and the generalized FR statistic by $\mathfrak{R}^{(ij)}_{n_1,n_j}$ that are computed as follows
\begin{itemize}
\item[1.] $\mathfrak{R}_{n_i n_j}:=\mathfrak{R}_{n_i n_j}(\BX)$ is the number of dichotomous edges in a Euclidean minimal spanning tree (MST) spanning the samples in the pairwise union of samples with labels $i$ and $j$ , $\BX^{(i)}\cup \BX^{(j)}\in \bbS^{(i)}\cup\bbS^{(j)}$, where $\BX^{(k)}=\{(\bx_i,y_i)\}_{i=1,Y_i=k}^n$. A dichotomous edge is an edge that connects a sample from class $i$ to a sample from class $j$. The pairwise FR statistic $\mathfrak{R}_{n_i n_j}$ for three classes is illustrated in Fig~\ref{fig:pairwiseghp_example}.  \\
\item[2.] $\mathfrak{R}^{(ij)}_{n_i,n_j}:=\mathfrak{R}^{(ij)}_{n_i,n_j}(\BX)$ is the number of dichotomous edges connecting a sample from class $i$ to a sample from class $j$ in the global MST constructed on all samples with all classes $1,2\ldots,m$ i.e. $\bigcup\limits_{k=1}^m \{(\bx_i,y_i)\}_{i=1,y_i=k}^n$ or  $\BX^{(1)}\cup\BX^{(2)}\cup\dots\cup\BX^{(m)}\in \bbS^{(1)}\cup \bbS^{(2)}\cup\dots \bbS^{(m)}$ where $\BX^{(k)}=\{(\bx_i,y_i)\}_{i=1,y_i=k}^n$, $k=1,\ldots,m$. Fig~\ref{fig:ghp_example} represents the generalized FR statistic for three classes. 
\end{itemize}
Using the theory in \cite{HP} and \cite{Berishaetal2016}, the estimator $\mathfrak{R}_{n_i,n_j}$ is a statistically consistent estimator of  the pairwise HP-bound for classifying class $i$ vs. class $j$. This yields an estimate of the bounds (11) on multi-class BER, requiring the construction of $m \choose 2$ MSTs spanning all distinct pairs of label classes. The next theorem implies that $\mathfrak{R}^{(ij)}_{n_i,n_j}$ can be used to estimate the tighter bound on multi-class BER given in Theorem \ref{thm.2} using only a single global MST. 
\begin{figure}[h]
\centering
\includegraphics[width=\columnwidth]{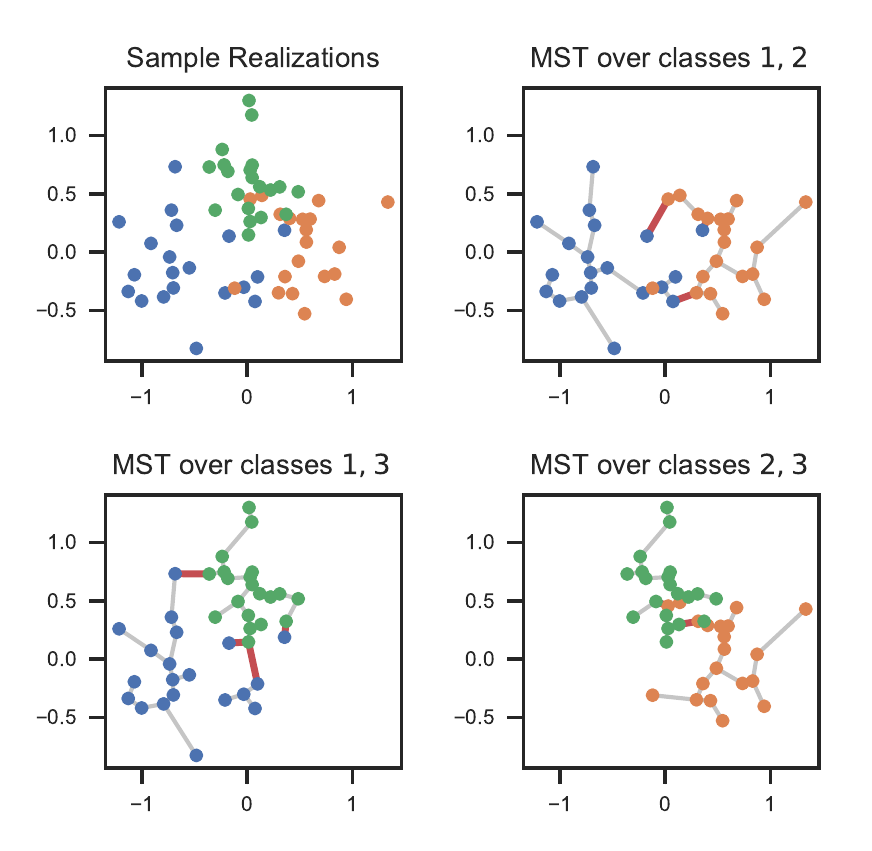}
\caption{Estimating $\mathfrak{R}_{n_i,n_j}$ for three classes. A set of $m(m-1)/2$ Euclidean MSTs are computed for each unordered pair of classes $\{(i,j)\}_{i>j}$, and then the dichotomous edges (in red) are counted to find $\mathfrak{R}_{n_i,n_j}$.}
\label{fig:pairwiseghp_example}
\end{figure}
\begin{figure}[h]
\centering
\includegraphics[width=3.2in] {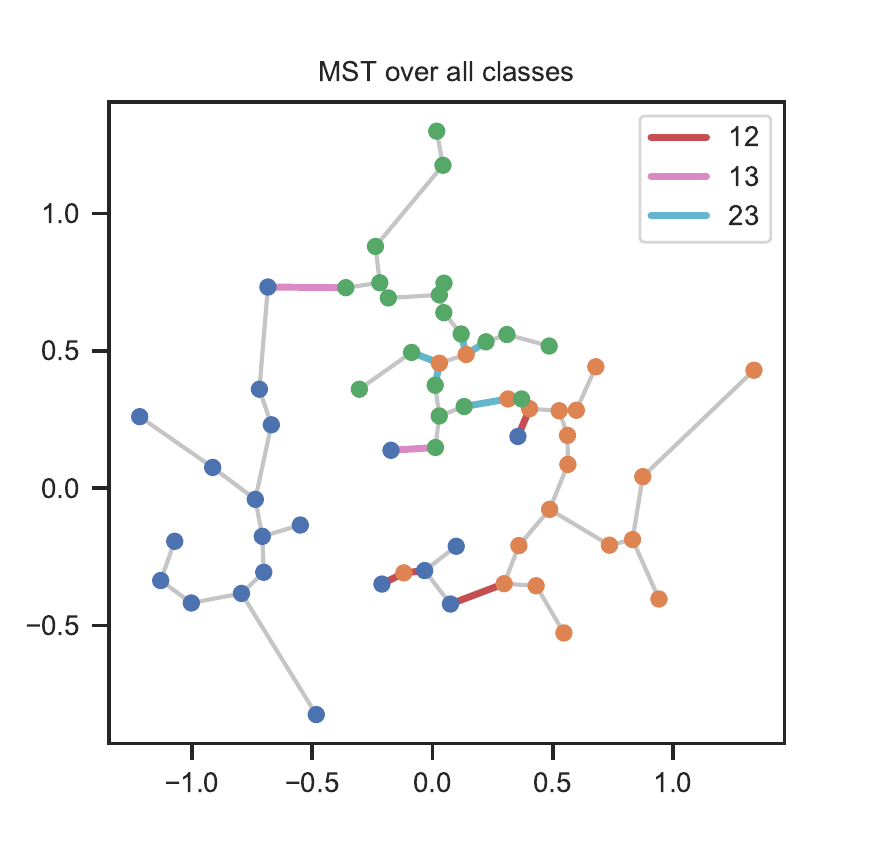}
\caption{Estimating $\mathfrak{R}^{(ij)}_{n_i,n_j}$ for three classes. A single MST is constructed over all classes $i=1,2,3$ of points. For each $(i, j)$, count the edges connecting points from classes $i$ and $j$. These edges are shown in 3 different colors each corresponding to the three types of pairs $(i,j)=(1,2), (2,3), (1,3)$.}
\label{fig:ghp_example}
\end{figure}

\begin{theorem}\label{thm:5}
{\it Let $\mathbf X= \{(\bx_i, y_i)\}_{i=1}^n$ be an i.i.d. $m$-class labeled sample and $n_k=\sum_{i=1}^n I(y_i=k)$, be the cardinality of samples with label $k$. For distinct classes $i,j$ let $i,j=1,\dots, m$, $n_i,n_j\rightarrow \infty$, $n\rightarrow \infty$ such that $n_i/n\rightarrow p_i$ and $n_j/n\rightarrow p_j$. Then
\begin{equation} \label{eq:V} \diy\frac{\mathfrak{R}^{(ij)}_{n_i,n_j}(\BX)}{2n}\diy\longrightarrow \delta^m_{ij}\;\; \;\; \hbox{(a.s.)}\end{equation}
}
\end{theorem}
The proof of Theorem \ref{thm:5} is given in Appendix G.

\section{Simulation Study}\label{sec.simulation}
\def\bbE{\mathbb{E}}
Here we illustrate the proposed method for learning bounds on Bayes error rate (BER).  Section~\ref{ssec:sim_bounds} focuses on numerical comparison of the upper bounds in~(\ref{multiclass.HP-bound}), (\ref{Upper-Lin}), (\ref{proposed.upbound}) and lower bounds in~(\ref{multiclass.HP-bound}), (\ref{JS:lowerbound}), (\ref{lowerbound.2}). Section~\ref{ssec:mst_sims} focuses on the empirical estimation of these bounds, including a comparison of runtime.

\begin{figure}[h]
\includegraphics[width=1.7in]{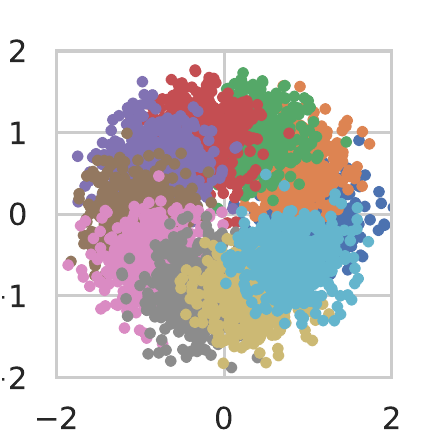}
\includegraphics[width=1.7in]{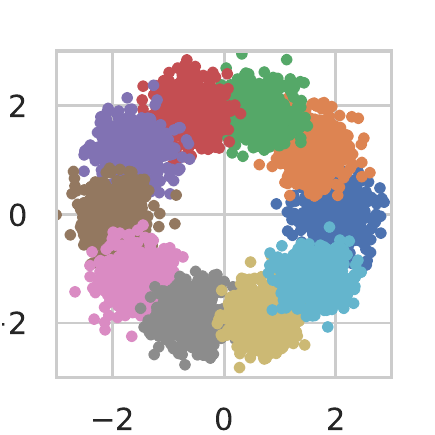}

\caption{Example of generated data for experiments. The data on the left has 10 classes whose means are arranged around a circle with mean parameter $\mu=1$. The data on the right has 10 classes, with mean parameter $\mu=2$. In both cases, $\sigma^2=0.1$, and both plots show a sample of 5000 data points.}
\label{fig:data_exs}
\end{figure}

\begin{figure}[h]
\hspace*{-0.9cm} 
\includegraphics[width=4.0in]{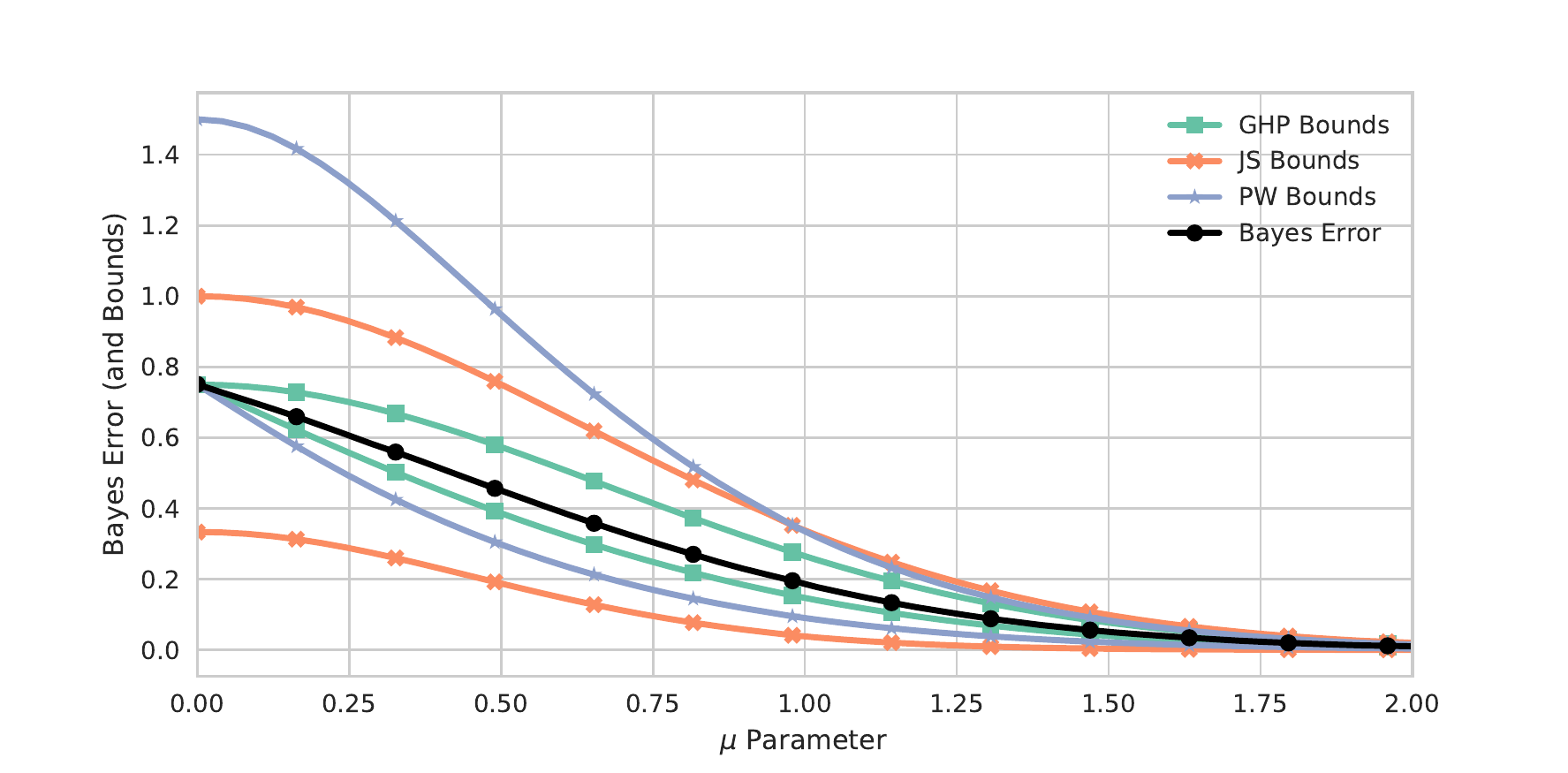}
\caption{Bounds on the Bayes error for $m=4$ and uniform priors. We note that even for a relatively small number of classes, the proposed new GHP upper and lower bound are much tighter than the competitors. For this experiment, $\sigma^2 = 0.3$.}
\label{fig:big_bound_fig}
\end{figure}

For each of the following simulations, data is generated in the following way: given $m$ classes with priors $p_1, p_2, \ldots, p_m$, the class conditional distributions are mean shifted bivariate normal: $f_i \sim \mathcal{N}(\mu_i, \sigma^2I)$. The means $\mu_i$ are arranged uniformly around the circumference of a circle of radius $\mu$:
\begin{equation*}
\mu_i = \left[\mu \cos \left( 2\pi \frac{i}{m} \right), \mu \sin \left( 2\pi \frac{i}{m} \right)\right].
\end{equation*}
Fig.~\ref{fig:data_exs} shows two examples for 5000 points and 10 classes and $\sigma^2=0.1$, with the left plot having mean parameter $\mu=0.7$, and the right plot setting $\mu=2$. Unless stated otherwise, the feature dimension is $d=2$.

Note that throughout this section, we compute Bayes error rate using Monte Carlo method. Specifically, we rewrite $\epsilon^m$ in terms of expectation $\mathbb{E}_\BX$ i.e.
\begin{equation}\label{MC} \diy \epsilon^{m}=\diy\bbE_\BX\Big[1-\max\limits_{i=1,\dots,m} p(i|\bx)\Big],\end{equation}
In addition, we know that $p(i|\bx)=f_i(\bx)p_i/f^m(\bx)$ and $f_i \sim \mathcal{N}(\mu_i, \sigma^2I)$. Therefore we generate sample from $f^m$ distribution and compute the RHS in (\ref{MC}) for the generated points, then we take the average to compute true BER. 

\subsection{Comparison of bounds}
\label{ssec:sim_bounds}

We first explore how the difficulty of the classification problem affects the bounds. Fig.~\ref{fig:big_bound_fig} shows upper and lower bounds of the Bayes error rate for each type of bound as a function of the mean parameter $\mu$. Here, the number of classes $m$ is 4. Note that when $\mu$ is smaller and the classes are poorly separated (creating a harder classification problem), both the Jensen-Shannon (JS) and pairwise (PW) upper bounds perform poorly and become trivial, exceeding one. However, for relatively small $m$, the pairwise lower bound remains fairly tight. The proposed GHP bounds are uniformly better than either the JS or PW bounds, as predicted by the theory. Further, note that the proposed bound is tight around the actual Bayes error rate (BER). Finally, as $\mu$ grows and the classes become well separated, the JS and PW bounds become tighter to the Bayes error. In light of Theorem 1, this makes sense for the pairwise bounds, as well separated classes cause the pairwise Henze-Penrose divergence and the GHP integral to become equivalent.



The difference between the bounds and the BER, called the tightness of the bound as a function of $m$ is shown in Fig.~\ref{fig:num_classes_upper} and Fig.~\ref{fig:num_classes_lower}  for upper bounds and lower bounds, respectively. Fig.~\ref{fig:num_classes_upper} highlights our proposed GHP bound's ability to stay close to the BER, even as the class size continues to increase. In comparison, both the JS and pairwise upper bounds continue to drift farther away from the Bayes error. Fig.~\ref{fig:num_classes_lower} shows a similar effectiveness in the proposed lower bound, although both the JS and pairwise bounds have better behavior than in Fig.~\ref{fig:num_classes_upper}, due to the lower bounds being guaranteed to be greater than or equal to 0.
\begin{figure}[h]
\centering
\includegraphics[width=3.0in]{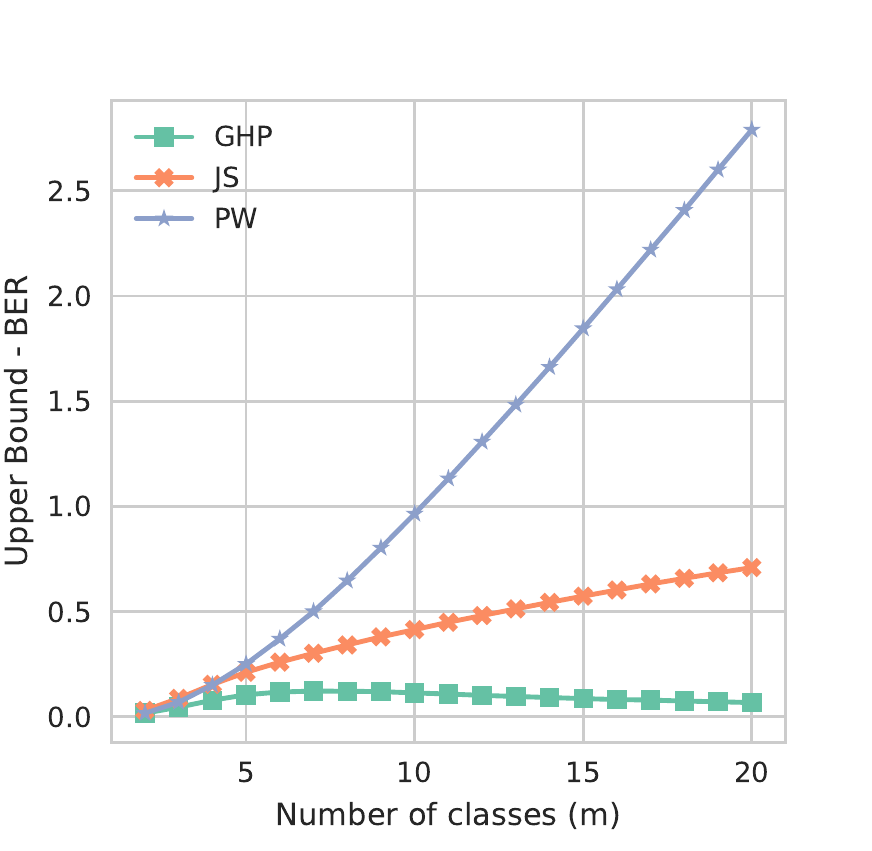}
\caption{Tightness of upper bound vs. $m$, where tightness is quantified as the difference between the upper bound and the true BER. This experiment was performed for $\mu = 1$. The pairwise upper bound quickly becomes useless as $m$ increases. The JS bound performs slightly better, but only our proposed GHP upper bound stays close to the Bayes error.}
\label{fig:num_classes_upper}
\end{figure}

\begin{figure}[h]
\vspace{-0.6cm}
\includegraphics[width=3.0in]{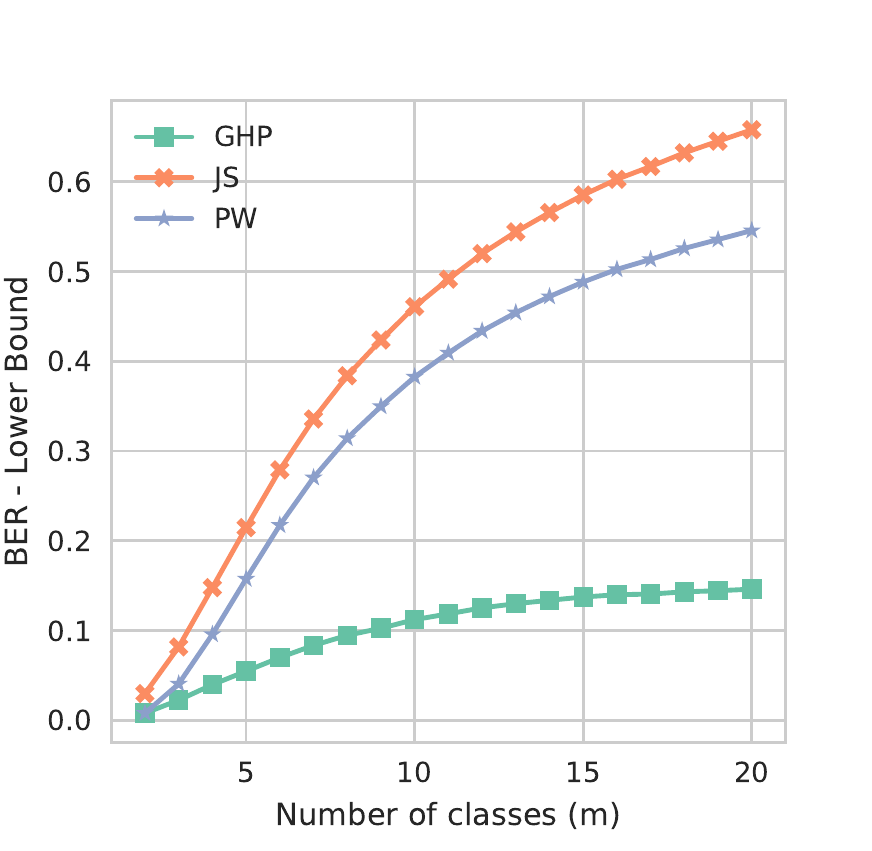}
\caption{Tightness of lower bound vs. $m$, defined as the difference between the BER and the lower bound. This experiment was performed for $\mu=1$. In the proposed GHP bound, there is a slight decrease in tightness as $m$ increases. However, it is much smaller in comparison with the pairwise and JS bound.}
\label{fig:num_classes_lower}
\end{figure}




\subsection{Statistical consistency and runtime}
\label{ssec:mst_sims}

This section illustrates the improvement in both statistical accuracy and runtime performance of the proposed generalized HP (GHP) approach as compared to the JS and pairwise HP (PW) methods of~\cite{Lin1991} and~\cite{BerishaICassp2016}.  

Fig.~\ref{fig:mst_convergence} Shows the MSE between the estimated and true upper bound as a function of the number of samples $n$, for different feature dimensions $d$. The behavior of the lower bound convergence has analogous behavior and is not shown. Note that as $d$ increases, the MSE grows, illustrating the well known curse of dimensionality for high dimensional datasets. 


%
\begin{figure}[h]
\includegraphics[width=3.4in]{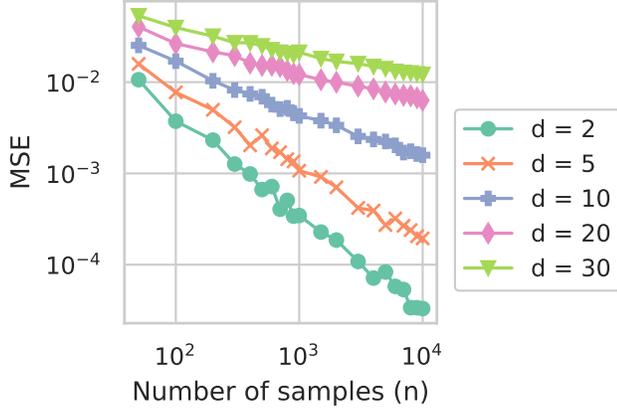}
\caption{Convergence in MSE of MST estimate of the proposed GHP upper bound to the true upper bound on BER. The simulation parameters were as in Fig.~\ref{fig:data_exs} for  $\mu=0.7$ and $\sigma^2=0.1$, and the results were averaged over 100 trials. For $d > 2$, $d - 2$ dimensions had zero mean with Gaussian noise having variance 0.1.} 
\label{fig:mst_convergence}
\end{figure}

Figs.~\ref{fig:runtime_m}~and~\ref{fig:runtime_n} show the relative runtime of the proposed method in comparison with the pairwise HP method. For each of these figures, we introduce a parameter $\gamma$, which controls the prior class imbalance in the problem. For a particular $\gamma$ and number of classes $m$, we create priors $p_1 = \gamma, p_2 = p_3 = \ldots = p_m = (1 - \gamma) / (m - 1)$.  For $\gamma=1/m$, all class probabilities are the same; there is no imbalance. A larger class imbalance will cause the pairwise estimation procedure to perform many large MST calculations, which significantly increases the runtime of the algorithm.
Fig.~\ref{fig:runtime_m} shows the relative runtime (pairwise - proposed method) as a function of $\gamma$, for different $m$, along with the ratio of tightness of GHP compared with PW for the upper bound of the BER. Here, we set $n = 10000, \mu=1, \sigma^2=0.3.$ Observe that for large number of classes and small class imbalance $\gamma$, the pairwise method is slightly faster  but, in this regime PW yields a useless bound that is overly loose - the proposed GHP bound is over 120 times tighter than the pairwise bound in this case. As $\gamma$ grows, we see significant relative speedup of the proposed GHP method relative to the other methods. From Fig. 11 it is evident that, while the PW bound has faster computation time in the regime of small $\gamma$ (graph in bottom panel), it is very loose in this small $\gamma$ regime  (graph in top panel).  
\begin{figure}[h]
\includegraphics[width=3.2in]{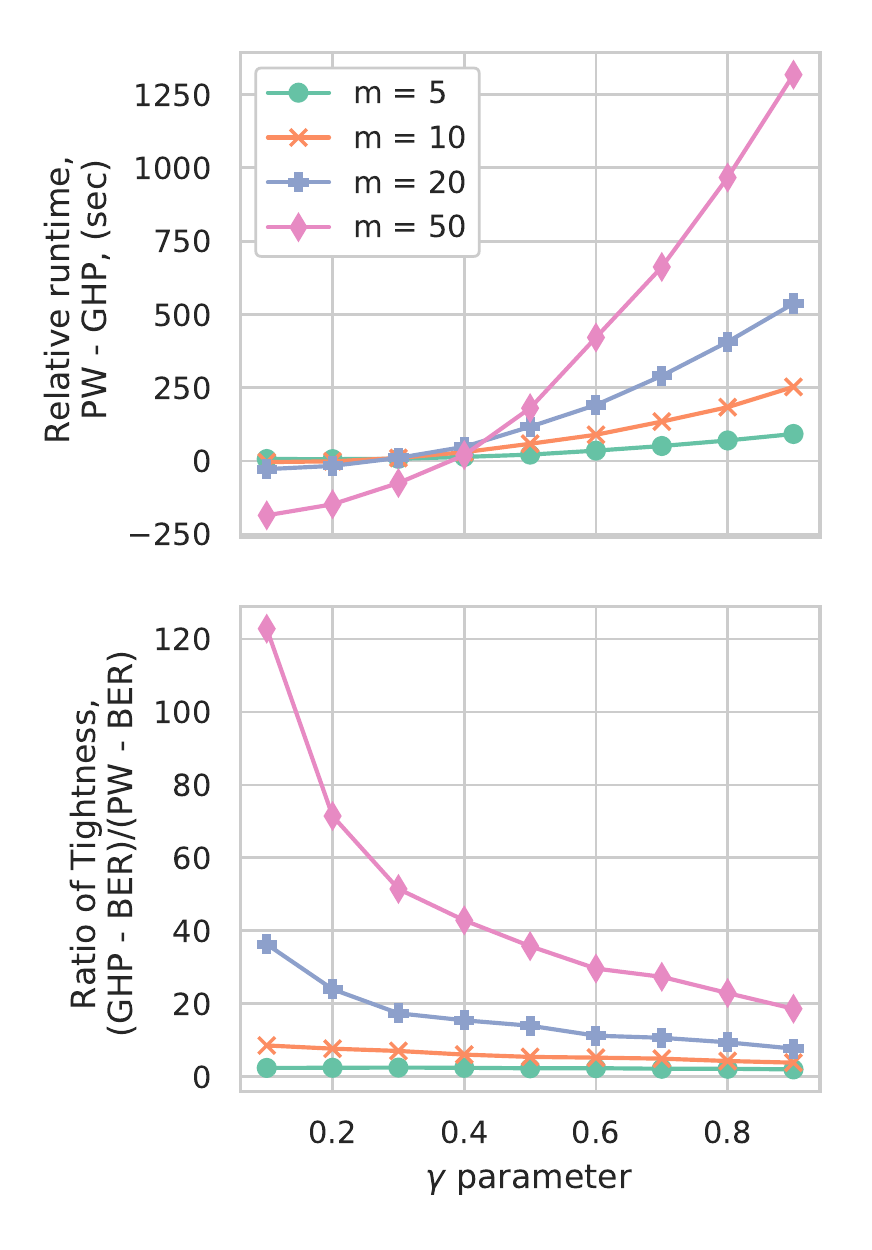}
\caption{Relative runtime of pairwise and proposed GHP algorithm vs. class imbalance parameter $\gamma$, and ratio of tightness of GHP compared with PW, where tightness is defined by the upper bound minus the BER. For large class imbalance (large $\gamma$), and large $m$, the proposed GHP method achieves significant speedup, while for small class imbalance, the PW bound becomes overly loose.}
\label{fig:runtime_m}
\end{figure}
Fig~\ref{fig:runtime_n} shows the relative runtime as a function of $\gamma$, for different sample sizes $n$, with $m=10, \mu=1,$ and $\sigma^2=0.3$. Similarly to Fig.~\ref{fig:runtime_m}, the greatest speedup occurs when $n$ and $\gamma$ are large. 

\begin{figure}[h]
\includegraphics[width=3.2in]{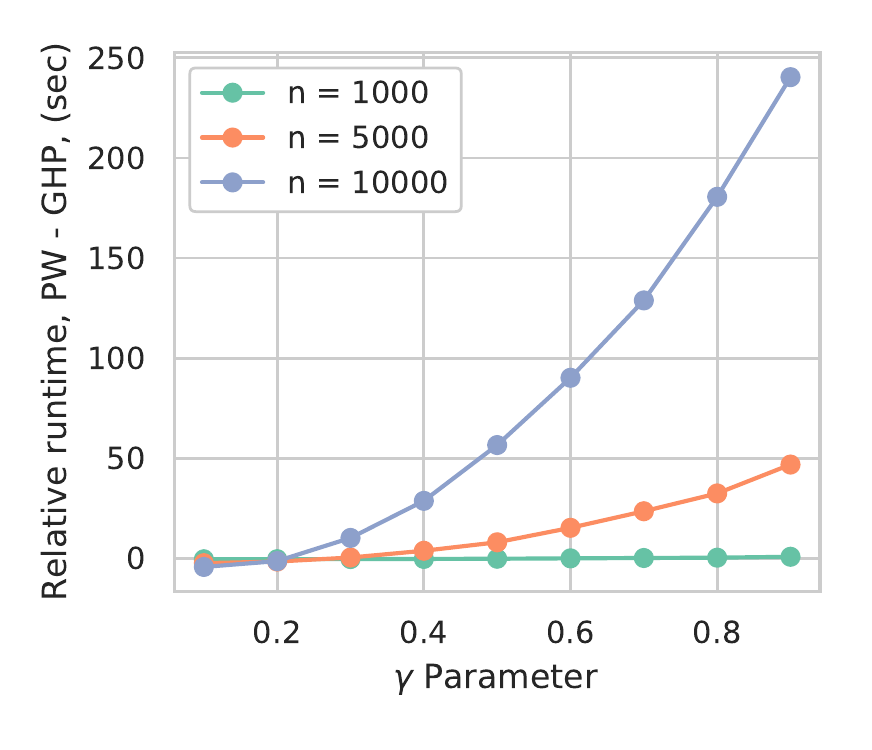}
\caption{Relative runtime of pairwise and proposed GHP algorithm vs. class imbalance parameter $\gamma$. For large $\gamma$, and large sample size $n$, the proposed method achieves significant speedup.}
\label{fig:runtime_n}
\end{figure}

\section{Real Data Experiments}
\label{Real-dataset}

\subsection{Using Bounds as Feature Extraction Quality}

We utilize our proposed bounds to explore feature generation for the MNIST dataset. The MNIST dataset consists of grey-scale thumbnails, 28 x 28 pixels, of hand-written digits 0 - 9. It consists of a training set of 60,000 examples, and a test set of 10,000 examples. The digits have been size-normalized and centered in a fixed-size image. MNIST has been well studied in the literature, and is known to have a low error-rate. To illustrate the utility of the proposed BER bound learning approach, we estimate the Bayes error rate bounds as a function of feature dimension. Specifically, we focus on PCA and a simple autoencoder. The validity of the proposed lower bound is demonstrated by comparison to the accuracy of three types of classifiers: the K-NN classifier, linear SVM, and a random forest classifier.


\begin {table*}[h]
\caption {Bounds on the Bayes Error and Classifier Test Error Rates for different feature sets. }
\begin{center}
\begin{tabular}{ |c||c|c|c|c|c|  }
\hline
 \multicolumn{6}{|c|}{Bayes Error Bounds and Error Rates for MNIST Feature Sets} \\
\hline
Features & lower bound & upper bound & Linear SVM & K-NN, K=3 & Rand. For. \\
 \hline
PCA-4           & 0.247 & 0.427 & 0.449 & 0.392 & 0.370 \\
PCA-8           & 0.070 & 0.135 & 0.241 & 0.107 & 0.129 \\
PCA-16          & 0.029 & 0.058 & 0.166 & 0.037 & 0.077 \\
PCA-32          & 0.020 & 0.040 & 0.113 & 0.026 & 0.073 \\
Autoencoder-4   & 0.290 & 0.486 & 0.662 & 0.442 & 0.412 \\
Autoencoder-8   & 0.097 & 0.184 & 0.317 & 0.144 & 0.155 \\
Autoencoder-16  & 0.041 & 0.082 & 0.213 & 0.058 & 0.099 \\
Autoencoder-32  & 0.026 & 0.052 & 0.144 & 0.032 & 0.086 \\
 \hline
\end{tabular}
\label{tab:real_data}
\end{center}

\end{table*}

The PCA results are shown in Fig.~\ref{fig:mnist_pca}. Plotted are the estimated lower bound for the BER and the test error rates of the 3-NN and Random Forest classifier versus the number of principal components used. As expected, the test errors of both classifiers are greater than the lower bound for Bayes error. Further, it is evident that no more than 20 latent dimensions are needed in order to minimize the lower bound, which is confirmed by the behavior of the test errors of the classifiers implemented.
\begin{figure}[h]
\vspace{-0.1cm}
\centering
\includegraphics[width=3.1in]{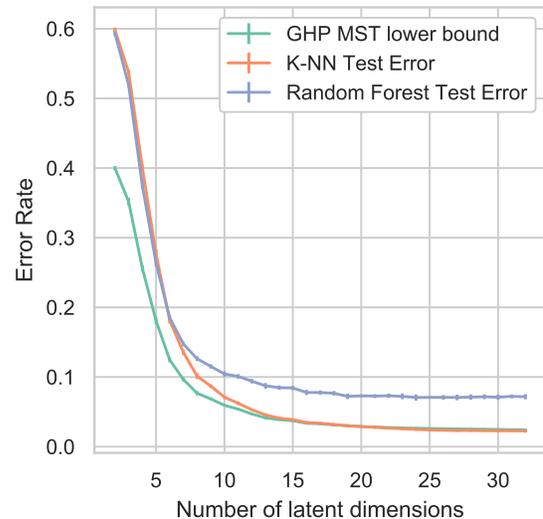}
\caption{Number of latent dimension for PCA vs. error rates and estimated lower bound. The k-NN classifier test error (orange curve) approaching the proposed lower bound (green curve) as the number of latent dimensions increases beyond 15, establishing that the k-NN comes close to achieving optimal performance.}
\label{fig:mnist_pca}
\end{figure}
Table~\ref{tab:real_data} shows Bayes error bounds and error rates for the MNIST feature sets. Autoencoder-X or PCA-X are feature sets that use X latent dimensions or X principal components, respectively. The autoencoder is a 1-layer autoencoder, and trained using Adagrad. Interestingly, we see that PCA-32 feature set outperforms autoencoder-32. More layers or a convolutional model could assist the autoencoder in achieving lower bounds and test error rates. 

\subsection{Application to CNN Network}
\label{ssec:cnn_network}

Another application of the proposed GHP bounds is to explore the layer-by-layer behavior of
neural networks. The cumulative effects on classification performance of the
layers can be analyzed by evaluating the proposed bounds at each intermediate
layer output. In order to compare similar numbers of features, auxiliary layers
--- in this case a ReLu and Softmax layer ---
are added after each intermediate layer that mimic the final two layers in the
network. Figure~\ref{fig:cnn_network} shows a graphical depiction of the architecture. We
utilize convolutional and max pooling layers, and test image datasets using this
framework.

\begin{figure}[h]
  \centering
   \includegraphics[width=\linewidth]{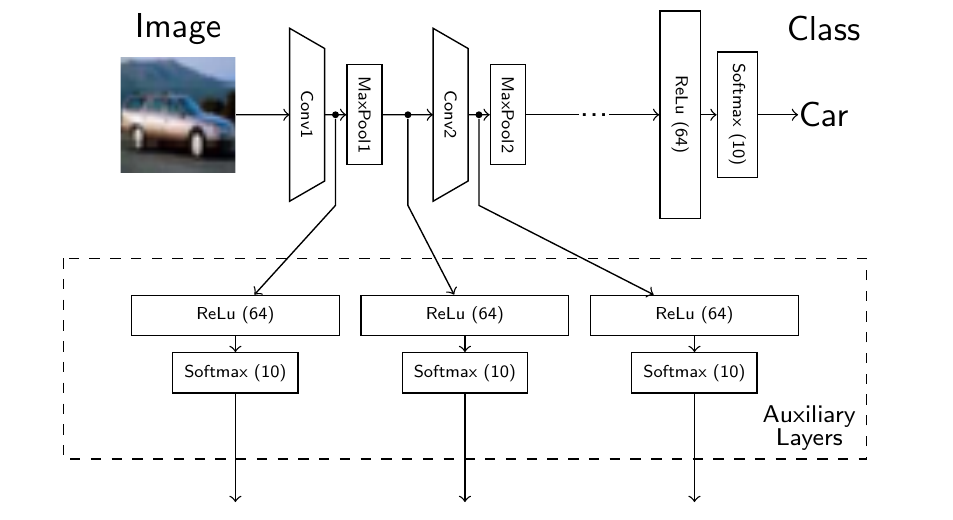}
  \caption{Convolutional neural network architecture. We utilize 3
    convolutional and max-pooling layers, along with two dense layers at the end, as well as
    in the auxiliary layers. The auxiliary layers are used to be able to compare the evolution of layer-by-layer  classification bounds for different CNN architectures using our proposed multiclass Bayes error estimator. Each convolutional layer has 16 channels.}
  \label{fig:cnn_network}
\end{figure}

The training proceeds as follows. The main pipeline is trained, and then these layers are frozen. Using these frozen layers, the auxiliary layers are then trained. Each part of the training used AdaGrad. Finally, feature output is taken at each endpoint for the auxiliary layers, and the bounds are subsequently calculated.  Each layer was initialized using standard random Normal variables, and the training was done for 25 runs of
the total dataset in question. Three datasets are utilized in this fashion: MNIST, CIFAR10, and CIFAR100. The results are shown in
Figure~\ref{fig:cnn_results}.
\begin{figure}[h]
  \centering
  \includegraphics[width=2.8in]{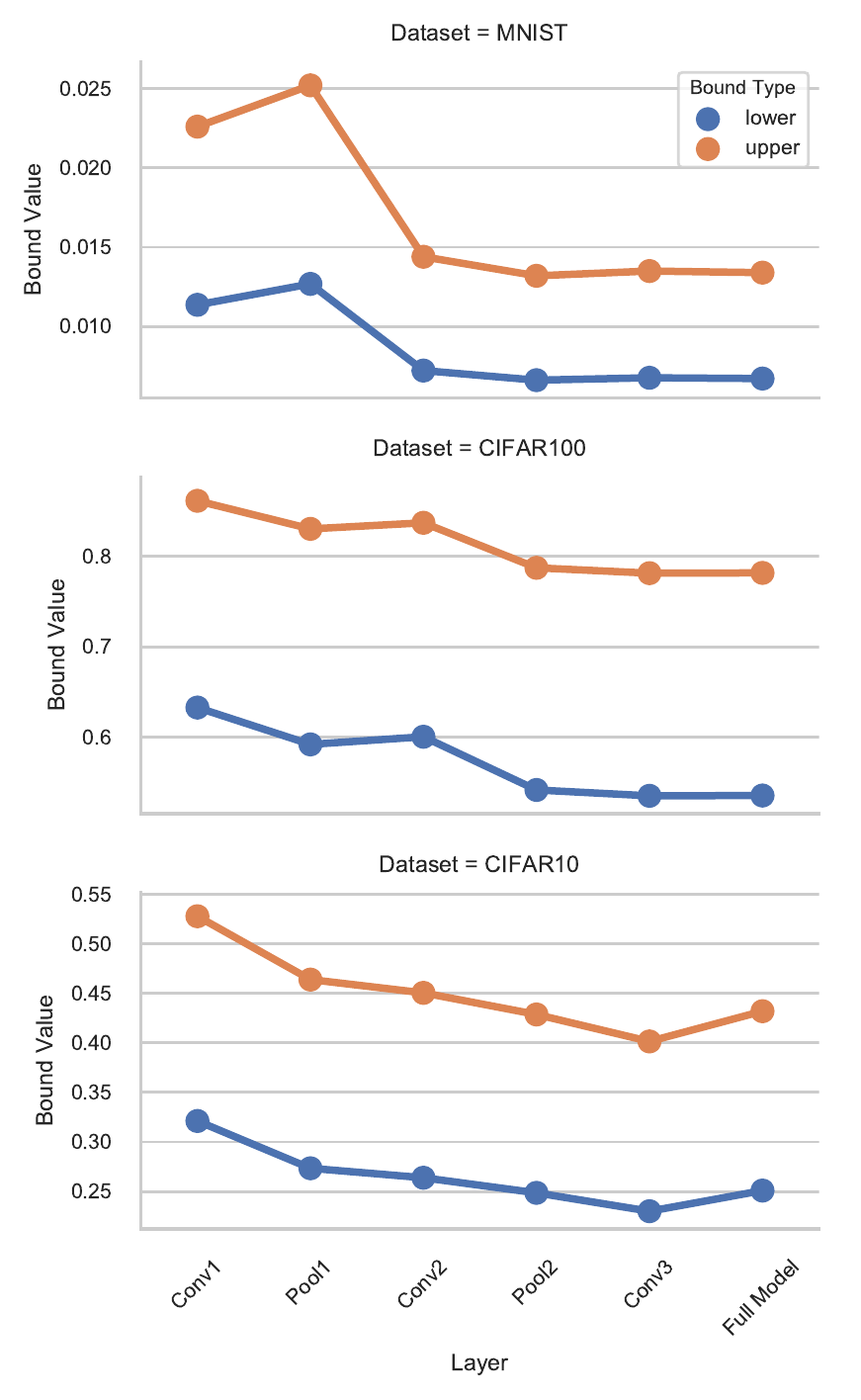}
  \caption{Performance bounds on the neural network, layer by layer. Most
    additional convolutional layer improves the performance of the overall
    network. Further, max-pooling does not significantly reduce the performance,
  as expected.}
  \label{fig:cnn_results}
\end{figure}
We note that, as expected, additional convolutional layers improves the
performance of the overall neural net. Further, we see that max-pooling does not
significantly affect the performance overall as measured by the proposed bounds, even though it is a downsampling operation. We further see that the bounds are able to distinguish between easy classification datasets such as MNIST, and more challenging problems such as CIFAR100.


\section{Conclusion}\label{sec:conclusion} 
In this paper, a new bound on the Bayes error rate of multiclass classification was introduced.  It was established by theory and simulation that the proposed bound is tighter than both the pairwise Henze-Penrose bound and the generalized Jenson-Shannon bound. Furthermore, a fast and efficient empirical estimator was presented that allows one to learn the bound from training data without the need for density estimation. The estimation method is based on the global minimal spanning tree that spans all labeled features over feature space, allowing for a more computationally tractable approach than the standard practice of summing estimates of pairwise BERs. 
The statistical complexity and statistical rates of convergence of the multiclass estimator would be useful for predicting the number of samples required to attain a specified accuracy of estimation of the proposed multiclass Bayes error bound. This is important area for future work.   
The proposed bound learning method was illustrated on the MNIST dataset.

\appendix
\section{Proofs of Main Theorems}\label{sec.proofs}
Here we first provide further discussion on (\ref{eq02:thm1}) and (\ref{eq03:thm1}) and then prove Theorems \ref{thm.2}-\ref{thm:5}. Throughout this section, we use notations $\delta_{ij}$ and $\delta_{ij}^m$ for FR and generalized FR test statistic, as defined in the paper. $D$ represents the HP divergence and $f^{(m)}(\bx)$ is the marginal distribution of random vector $\BX$; $\bbE$ stands for expectation. 
\subsection{Discussion on (\ref{eq02:thm1}) and (\ref{eq03:thm1})}
The relation (\ref{eq02:thm1}) can be easily derived. Here we provide the proof of relation (\ref{eq03:thm1}). It can be seen that there exists a constant $C_1$ depending on the $p_i$ and $p_j$ such that for every $f_i$ and $f_j$ 
\begin{equation}\label{equ.thm1}
f_i(\bx)f_j(\bx)\leq C_1\left(p_if_i(\bx)+p_jf_j(\bx)\right)^2.
\end{equation}
Set 
\begin{equation}
g_{ij}(\bx):=\diy\left(p_if_i(\bx)+p_jf_j(\bx)\right)\diy\sum\limits_{k\neq i,j}p_kf_k(\bx)\big/f^{(m)}(\bx). 
\end{equation}
The inequality (\ref{equ.thm1}) is equivalent to 
\begin{equation}
0\leq f_i(\bx)f_j(\bx)\left(\diy\frac{1}{p_if_i(\bx)+p_jf_j(\bx)}-\frac{1}{f^{(m)}(\bx)}\right)\leq C_1 \; g_{ij}(\bx).
\end{equation}
Therefore 
\begin{equation}\label{Ineq:thm1.proof}\begin{array}{l}
\diy\int\diy\frac{f_i(\bx)f_j(\bx)}{p_if_i(\bx)+p_jf_j(\bx)}\rd\bx\leq {\it \diy\int{\it \diy\frac{f_i(\bx)f_j(\bx)}{f^{(m)}(\bx)}\rd\bx}
+C_1\Big(\diy\int g_{ij}(\bx)\;\rd\bx\Big)}
\end{array}\end{equation}
On the other hand, we have
\begin{equation}\begin{array}{l}
D\Big(\widetilde{p}_{ij}f_i+\widetilde{p}_{ji}f_j,\diy\sum\limits_{k\neq i,j}\widetilde{p}_k^{ij}f_k\Big)=\\
\qquad\qquad1-\diy\frac{1}{(p_i+p_j)\diy\sum\limits_{r\neq i,j}p_r}\diy\int g_{ij}(\bx)\;\rd\bx,
\end{array}\end{equation}
where $\widetilde{p}_{ij}$ and $\widetilde{p}_{ji}$ are as before and $\widetilde{p}_k^{ij}=p_k\big/\diy\sum\limits_{r\neq i,j}p_r$. Hence, 
\begin{equation}
\int g_{ij}(\bx)\;\rd\bx=C_2{\it \Big\{1-D\Big(\widetilde{p}_{ij}f_i+\widetilde{p}_{ji}f_j,\diy\sum\limits_{k\neq i,j}\widetilde{p}_k^{ij}f_k\Big)\Big\}}.
\end{equation}
where $C_2$ is a constant depending on priors $p_1,p_2,\dots p_m$. This together with (\ref{Ineq:thm1.proof}) implies that there exists a constant $C$ depending only on priors $p_1,p_2,\dots,p_m$ such that 
\begin{equation} \begin{array}{l}
\diy\int\diy\frac{f_i(\bx)f_j(\bx)}{p_if_i(\bx)+p_jf_j(\bx)}\rd\bx\leq{\it \diy\int\diy\frac{f_i(\bx)f_j(\bx)}{f^{(m)}(\bx)}\rd\bx}\\
\\
\qquad \qquad +C\Big(1-D\Big(\widetilde{p}_{ij}f_i+\widetilde{p}_{ji}f_j,\diy\sum\limits_{k\neq i,j}\widetilde{p}_k^{ij}f_k\Big)\Big),
\end{array}\end{equation}

By recalling ${\rm HP}_{ij}$ (\ref{Def:HP}) and ${\rm GHP}^m_{ij} $ (\ref{Def:GHP}) we conclude the result. 
\subsection{Theorem \ref{thm.2}}
To derive the inequality in (\ref{proposed.upbound}), first we need to prove the following lemma:
\begin{lemma}\label{lem.1}
Let $a_1,a_2,\dots,a_m$ be a probability distribution on $m$ classes so that $\diy\sum\limits_{i=1}^m a_i=1$. Then 
\begin{equation}
1-\max\limits_i a_i\leq 2\diy\sum\limits_{i=1}^{m-1}\sum\limits_{j=i+1}^{m} a_i a_j. 
\end{equation}
\end{lemma}
\begin{IEEEproof}
Assume, without loss of generality, that the $a_i$ have been reordered in such a way that $a_m$ is the largest. So it is sufficient to prove that
\begin{equation}
1-a_m\leq 2\diy\sum\limits_{i=1}^{m-1}\sum\limits_{j=i+1}^m a_i a_j. 
\end{equation}
Since $\diy\sum\limits_{i=1}^ma_i=1$ then
$$1-a_m=\diy\sum\limits_{i=1}^{m-1}a_i=\diy\sum\limits_{i=1}^{m-1}\sum\limits_{j=1}^m a_i a_j.$$
Therefore we need to show that
\begin{equation}\label{eq1:lemma1}
\diy\sum\limits_{i=1}^{m-1}\sum\limits_{j=1}^m a_i a_j\leq 2\diy\sum\limits_{i=1}^{m-1}\sum\limits_{j=i+1}^m a_i a_j.
\end{equation}
The LHS in (\ref{eq1:lemma1}) is 
\begin{equation}\label{eq2:lemma1}
\diy\sum\limits_{i=1}^{m-1}\sum\limits_{j=1}^m a_i a_j=2 \diy\sum\limits_{i=1}^{m-2}\sum\limits_{j=i+1}^{m-1}a_i a_j+\sum\limits_{i=1}^{m-1}a_ia_m+\sum\limits_{i=1}^{m-1}a_i^2.
\end{equation}
And the RHS in (\ref{eq1:lemma1}) is written as
\begin{equation}\label{eq3:lemma1}
2\diy\sum\limits_{i=1}^{m-1}\sum\limits_{j=i+1}^m a_i a_j=2 \diy\sum\limits_{i=1}^{m-2}\sum\limits_{j=i+1}^{m-1}a_i a_j+2\diy\sum\limits_{i=1}^{m-1}a_i a_m.
\end{equation}
Recalling our assumption that $a_m$ is the largest we have 
\begin{equation}
\diy\sum\limits_{i=1}^{m-1}a_i^2\leq \diy\sum\limits_{i=1}^{m-1}a_i a_m.
\end{equation}
This implies that $(\ref{eq2:lemma1})\leq (\ref{eq3:lemma1})$. This concludes (\ref{eq1:lemma1}) and proves our Lemma. 
\end{IEEEproof}

Going back to prove upper bound (\ref{proposed.upbound}) in Theorem \ref{thm.2}, let $p_1f_1(\bx),p_2f_2(\bx),\dots,p_mf_m(\bx)$ be joint probabilities of $\bx$ and $i$. And denote $p(i|\bx):=P(y=i|\bx)$ where variable $y\in \{1,2,\dots,m\}$ is class label with priors $p_i$. The BER for $m$ classes is given by
\begin{equation}\begin{array}{cl}
\epsilon^{m}=1-\diy\int \max\big\{p_1f_1(\bx),\dots,p_mf_m(\bx)\big\}\rd\bx\\
=\bbE_\BX\Big[1-\max\limits_{i=1,\dots,m} p(i|\bx)\Big],
\end{array}\end{equation}
Moreover the marginal density for random vector $\BX$ is 
$$f_{\BX}(\bx)=\diy\sum\limits_{i=1}^m p_i f_i(\bx)=f^{(m)}(\bx),$$
And 
\begin{equation}\begin{array}{l}
\diy{\int}\frac{p_ip_jf_i(\bx)f_j(\bx)}{f^{(m)}(\bx)}{\rm d}\bx=\diy{\int}\left(\frac{ p_i f_i(\bx)}{f^{m}(\bx)}\right)\left(\frac{p_jf_j(\bx)}{f^{(m)}(\bx)}\right) f^{(m)}(\bx)\rd\bx\\
\\
\qquad\quad=\diy{\int}p(i|\bx)p(j|\bx) f^{(m)}(\bx){\rm d}\bx=\bbE_{\BX}\left[p(i|\BX)p(j|\BX) \right]. 
\end{array}\end{equation}
Therefore (\ref{proposed.upbound}) turns into  the following claim:
\begin{equation}\label{eq4.1.thm2}
\diy \bbE_\BX\Big[1-\max\limits_{i=1,\dots,m} p(i|\bx)\Big]\leq \bbE_{\BX}\left[2\diy\sum\limits_{i=1}^{m-1}\sum\limits_{j=i+1}^m p(i|\BX)p(j|\BX) \right].
\end{equation}
We know that $\diy\sum\limits_{i=1}^m p(i|\bx)=1$. Using Lemma \ref{lem.1} where $a_i$ represents $p(i|\bx)$ we have
\begin{equation}
\diy 1-\max\limits_{i=1,\dots,m} p(i|\bx)\leq 2\diy\sum\limits_{i=1}^{m-1}\sum\limits_{j=i+1}^m p(i|\BX)p(j|\BX).
\end{equation}
Hence, we prove the inequality (\ref{eq4.1.thm2}) and consequently our claim (\ref{proposed.upbound}). 

Next we prove the lower bound (\ref{lowerbound.2}). The following lemma is required:
\begin{lemma}
For all $a_1, a_2, ..., a_m$ such that $\diy\sum_{i=1}^m a_i = 1$, we have the following:
\begin{equation}\label{lemma.lowerbound}
\frac{m-1}{m}\left[ 1 - \left(1 - 2\frac{m}{m-1} \sum_{i=1}^{m-1} \sum_{j=i+1}^{m} a_i a_j\right)^{1/2}\right] \le 1 - \max_i a_i.
\end{equation}
\end{lemma}
\begin{IEEEproof}
After some algebra, we rewrite the inequality in the following form:

$$
m (\max_i a_i)^2 - 2 \max_i a_i \le m - 2 - (m - 1)b,
$$

where $b = \diy 2\sum_{i=1}^{m-1}\sum_{j=i+1}^{m}a_i a_j.$ Without loss of generality, we can assume that the $a_i$s are ordered, so that $a_m$ is the largest. Then we have that $\max_i a_i = 1 -\diy \sum_{i=1}^{m-1} a_i.$

Using this equality on the left side, expanding the square, and subtracting $m-2$ from both sides, we have:

\begin{equation}
m (\sum_{i=1}^{m-1}a_i)^2 - (2m - 1) \sum_{i=1}^{m-1}a_i \le -(m - 1)b.
\end{equation}

Expanding terms once again:
\begin{equation}
\sum_{i=1}^{m-1}\sum_{j=i+1}^m a_i a_j=\sum_{i=1}^{m-2}\sum_{j=i+1}^{m-1} a_i a_j+\sum_{i=1}^{m-1}a_i a_m,
\end{equation}
 and collecting like terms:

\begin{equation}\begin{array}{l}
\diy m \sum_{i=1}^{m-1}a_i^2 + (4m - 2) \sum_{i=1}^{m-2}\sum_{j=i+1}^{m-1}a_i a_j - (2m - 1) \sum_{i=1}^{m-1}a_i \\
\\
\qquad \qquad \le -\diy 2(m -1) \sum_{i=1}^{m-1} a_i a_m
\end{array}\end{equation}

We note, that since $\sum_{i=1}^m a_i = 1$, we have the following:

$$
\sum_{i=1}^{m-1}a_i = \sum_{i=1}^{m-1}\sum_{j=1}^m a_i a_j = 2 \sum_{i=1}^{m-2}\sum_{j=i+1}^{m-1} a_i a_j + \sum_{i=1}^{m-1}a_i a_m + \sum_{i=1}^{m -1}a_i^2.
$$

Plugging in once more:

$$
(1 - m)\sum_{i=1}^{m -1}a_i^2 - (2m - 1) \sum_{i=1}^{m - 1}a_i a_m \le  -2(m -1) \sum_{i=1}^{m - 1}a_i a_m, 
$$

or equivalently:
$$
(1 - m)\sum_{i=1}^{m -1}a_i^2 - \sum_{i=1}^{m - 1}a_i a_m \le 0.
$$

Note that since $a_m = \max_i a_i$, $\diy \sum_{i=1}^{m-1} a_i a_m \ge \diy \sum_{i=1}^{m-1} a_i^2$, so that

$$
(1 - m)\sum_{i=1}^{m -1}a_i^2 - \sum_{i=1}^{m - 1}a_i a_m \le -m \sum_{i=1}^{m -1}a_i^2 \le 0,
$$
since $\diy\sum_{i=1}^{m -1}a_i^2 \ge 0$.
\end{IEEEproof}
Now to prove (\ref{lowerbound.2}), let $p_1f_1(\bx),p_2f_2(\bx),\dots,p_mf_m(\bx)$ be joint probabilities of $\bx$ and $i$. And denote $p(i|\bx):=P(y=i|\bx)$ where variable $y\in \{1,2,\dots,m\}$ is class label with priors $p_i$. By taking the expectation from both sides of (\ref{lemma.lowerbound}) when $a_i=p(i|\bx)$, we have
\begin{equation}\label{eq1.lemma.LB}\begin{array}{l}
\diy \bbE_\BX[1-\max_{i} p(i|\bx)]\\
\quad\geq \diy\frac{m-1}{m}\left[ 1 - \bbE_\BX\left(1 - 2\frac{m}{m-1} \sum_{i=1}^{m-1} \sum_{j=i+1}^{m} p(i|\bx) p(j|\bx)\right)^{1/2}\right],
\end{array}\end{equation}
Further, since $\phi(\bx)=\sqrt{\bx}$ is a concave function, by applying Jensen inequality the RHS in (\ref{eq1.lemma.LB}) is lower bounded by 
\begin{equation}
\diy\frac{m-1}{m}\left[ 1 - \left(1 - 2\frac{m}{m-1} \bbE_\BX\left[\sum_{i=1}^{m-1} \sum_{j=i+1}^{m} p(i|\bx) p(j|\bx)\right]\right)^{1/2}\right],
\end{equation}
And we know that $$\bbE_{\BX}\left[p(i|\bx)p(j|\bx)\right]=\delta^m_{ij},$$ and 
$$\bbE_\BX[1-\max_{i} p(i|\bx)]=\ep^m,$$
then this proves our proposed lower bound in (\ref{lowerbound.2}).

\subsection{Theorem \ref{GHPandJS.upbound}}\label{GHPandJS.UB}
To derive (\ref{Ineq:tighterJS}), the following lemma is required to be proved:
\begin{lemma}\label{lem:GHPandJS.upbound}
Let $a_1,a_2,\dots,a_m$ be probability distributions on $m$ classes so $\diy\sum\limits_{i=1}^m a_i=1$. Then, for $m\geq 3$ and log basis $2$, we have
\begin{equation}\label{eq1:GHPandJS.upbound}
2\diy\sum_{i=1}^{m-1}\sum_{j=i+1}^m a_i a_j\leq -\diy\frac{1}{2} \sum_{i=1}^m a_i\log a_i. 
\end{equation}
\end{lemma}
\begin{IEEEproof}
The claim in (\ref{eq1:GHPandJS.upbound}) can be rewritten as
\begin{equation}\label{eq2:GHPandJS.upbound}
4\diy\sum_{i=1}^{m-1}\sum_{j=i+1}^m a_i a_j\leq \diy\sum_{i=1}^m a_i\log \diy\frac{1}{a_i}, 
\end{equation}
where $0\leq a_i\leq 1$. In addition we have 
\begin{equation}\label{eq3:lemma02}
4\diy\sum\limits_{i=1}^{m-1}\sum\limits_{j=i+1}^m a_i a_j=4 \diy\sum\limits_{i=1}^{m-2}\sum\limits_{j=i+1}^{m-1}a_i a_j+4\diy\sum\limits_{i=1}^{m-1}a_i a_m,
\end{equation}
and 
\begin{equation}\label{eq4:lemma02}
\sum_{i=1}^{m-1}a_i - \sum_{i=1}^{m-1}a_i a_m - \sum_{i=1}^{m -1}a_i^2= 2 \sum_{i=1}^{m-2}\sum_{j=i+1}^{m-1} a_i a_j .
\end{equation}
Combining (\ref{eq3:lemma02}) and (\ref{eq4:lemma02}), we have 
\begin{equation}\label{eq1:UB.GHP.JS}\begin{array}{ccl}
4\diy\sum\limits_{i=1}^{m-1}\sum\limits_{j=i+1}^m a_i a_j&=&\diy2\sum_{i=1}^{m-1}a_i + 2\sum_{i=1}^{m-1}a_i a_m - 2\sum_{i=1}^{m -1}a_i^2\\
&=&\diy 2(1-a_m^2)- 2\sum_{i=1}^{m -1}a_i^2.
\end{array}\end{equation}
Hence we need to show that 
\begin{equation}
\diy 2(1-a_m^2)- 2\sum_{i=1}^{m -1}a_i^2\leq \diy\sum_{i=1}^m a_i\log \diy\frac{1}{a_i}.
\end{equation}
Equivalently 
\begin{equation}
2-2\sum_{i=1}^m a_i^2\leq \diy\sum_{i=1}^m a_i\log\diy\frac{1}{a_i}.
\end{equation}
Or 
\begin{equation}\label{g.function}
g(m):=\sum_{i=1}^m a_i(2-2a_i +\log a_i)\leq 0.
\end{equation}
Since for $a_i\leq 1/2$ the function $a_i\big(2-2a_i+\log a_i\big)$ is negative and we know that $\diy\sum_{i=1}^m a_i=1$, therefore $g(m)$ is a decreasing function in $m$ i.e. $g(m)\leq g(3)$ for $m\geq 3$. And it can be easily checked that $g(3)\leq 0$. Hence the proof is completed.  

\end{IEEEproof}

Now, Following arguments in \cite{Lin1991}, one can check that
\begin{equation}
\diy\frac{1}{2}\big(H(p)-JS(f_1,f_2,\dots,f_m)\big)=-\diy\frac{1}{2}\bbE_{\BX}\left[\sum_{i=1}^m p(i|\BX)\log p(i|\BX)\right]. 
\end{equation}
Further, in Theorem \ref{thm.2}, we derived 
\begin{equation}
 2\sum\limits_{i=1}^{m-1}\sum\limits_{j=i+1}^m \delta^m_{ij}=\bbE_{\BX}\left[2\diy\sum\limits_{i=1}^{m-1}\sum\limits_{j=i+1}^m p(i|\BX)p(j|\BX)\right],
\end{equation}
such that $\diy\sum\limits_{i=1}^m p(i|\bx)=1$. Using Lemma \ref{lem:GHPandJS.upbound}, where again $a_i=p(i|\bx)$, we have 
\begin{equation}\label{eq3:GHPandJS.upbound}
2\diy\sum\limits_{i=1}^{m-1}\sum\limits_{j=i+1}^m p(i|\BX)p(j|\BX)\leq -\diy\frac{1}{2}\sum_{i=1}^m p(i|\BX)\log p(i|\BX).
\end{equation}
Taking expectation from both sides of (\ref{eq3:GHPandJS.upbound}) proves our claim in (\ref{Ineq:tighterJS}). 

Next, we prove the lower bound in (\ref{lowerbound.1.tight}). Similar to Appendices B and C let $p(i|\bx)$ be the posterior probabilities. Therefore we can rewrite  (\ref{lowerbound.1.tight}) in terms of $p(i|\bx)$ as
\begin{equation}\label{eq1:LB.GHP.JS}\begin{array}{l}
\diy\frac{m-1}{m}\left[ 1 - \left(1 - 2\frac{m}{m-1} \bbE_\BX\left[\sum_{i=1}^{m-1} \sum_{j=i+1}^{m} p(i|\bx) p(j|\bx)\right]\right)^{1/2}\right]\\
\\
\qquad \quad\geq \diy\frac{1}{4(m-1)}\diy\left(\bbE_{\BX}\left[\sum_{i=1}^m p(i|\bx)\log p(i|\bx)\right]\right)^2.
\end{array}\end{equation}
Analogous to other proofs, let $a_i=p(i|\bx)$ and to shorten the formula set 
$$ A(\bx)=1 - 2\frac{m}{m-1}\left[\sum_{i=1}^{m-1} \sum_{j=i+1}^{m} a_i a_j\right], $$
therefore (\ref{eq1:LB.GHP.JS}) can be rewritten as 
\begin{equation}
    \diy\frac{m-1}{m}\left[ 1 -\sqrt{\bbE_\BX[A(\BX)]}\right]\geq \diy\frac{1}{4(m-1)}\left(\bbE_\BX\left[\sum\limits_{i=1}^m a_i\log a_i\right]\right)^2.
\end{equation}
Equivalently 
\begin{equation}\label{eq2:LB.GHP.JS}
    \diy\left[ 1 -\sqrt{\bbE_\BX[A(\BX)]}\right]\geq \diy\frac{m}{4(m-1)^2}\left(\bbE_\BX\left[\sum\limits_{i=1}^m a_i\log a_i\right]\right)^2.
\end{equation}
Multiple the both sides of (\ref{eq1:LB.GHP.JS}) in $1 +\sqrt{\bbE_\BX[A(\BX)]}$:
\begin{equation}\label{eq2:LB.GHP.JS}\begin{array}{l}
    \diy\left[ 1 -\bbE_\BX[A(\BX)]\right]\\
    \qquad \geq \diy\frac{m}{4(m-1)^2}\left(\bbE_\BX\left[\sum\limits_{i=1}^m a_i\log a_i\right]\right)^2 \left(1 +\sqrt{\bbE_\BX[A(\BX)]}\right).
\end{array}\end{equation}
And we have 
$$ 1 -\bbE_\BX[A(\BX)]=2\diy\frac{m}{m-1}\bbE_{\BX}\left[\sum_{i=1}^{m-1} \sum_{j=i+1}^{m} a_i a_j\right].$$
And since $\sqrt{\bbE_\BX[A(\BX)]}\leq 1$ then $1 +\sqrt{\bbE_\BX[A(\BX)]}\leq 2$, so it is sufficient to prove that 
\begin{equation}
    \bbE_{\BX}\left[\sum_{i=1}^{m-1} \sum_{j=i+1}^{m} a_i a_j\right]\geq \diy\frac{1}{4(m-1)} \left(\bbE_\BX\left[\sum\limits_{i=1}^m a_i\log a_i\right]\right)^2.
\end{equation}
On the other hand we know that by using Jensen inequality
$$ \left(\bbE_\BX\left[\sum\limits_{i=1}^m a_i\log a_i\right]\right)^2 \leq \bbE_\BX\left[\sum\limits_{i=1}^m a_i\log a_i\right]^2, 
$$
so we only need to show that
\begin{equation}
    \sum_{i=1}^{m-1} \sum_{j=i+1}^{m} a_i a_j\geq \diy\frac{1}{4(m-1)} \left[\sum\limits_{i=1}^m a_i\log a_i\right]^2.
\end{equation}
Or 
\begin{equation}
    4(m-1)\sum_{i=1}^{m-1} \sum_{j=i+1}^{m} a_i a_j\geq \diy \left[\sum\limits_{i=1}^m a_i\log a_i\right]^2.
\end{equation}
Recalling (\ref{eq1:UB.GHP.JS}) in Appendix \ref{GHPandJS.UB} this is equivalent to 
\begin{equation}
2(m-1)\left(1-\sum\limits_{i=1}^m a_i^2\right)\geq \left[\sum\limits_{i=1}^m a_i\log a_i\right]^2.
\end{equation}
Now let $g(m)$ be
$$ 2(m-1)\left(1-\sum\limits_{i=1}^m a_i^2\right)- \left[\sum\limits_{i=1}^m a_i\log a_i\right]^2,$$
this is non-negative when $m=3$, $g(3)\geq 0$. In addition $g$ is an increasing function in $m$ i.e. $g(m)\geq g(3)$. Therefore following similar arguments as showing (\ref{g.function}) the proof of (\ref{lowerbound.1.tight}) is completed. 



\subsection{Theorem \ref{thm.lower.tight.2}}
Recalling the pairwise bound (\ref{multiclass.HP-bound}), the multi-class classification Bayes error HP bound is given as
\begin{equation}\label{multiclass-HP-bound}
\ep^m\leq 2\diy\sum\limits_{i=1}^{m-1}\sum\limits_{j=i+1}^m \delta_{ij}.
\end{equation}
Since $\delta_{ij}^m\leq \delta_{ij}$, our proposed bound (\ref{proposed.upbound}) is tighter than (\ref{multiclass-HP-bound}). This implies (\ref{upper.toghterPW}). 

To derive (\ref{lowerbound.2.tight}), let us first focus on $u_{\tilde{p}_{ij}}$:
\begin{equation}\begin{array}{ccl}
   u_{\tilde{p}_{ij}}&=&\diy 4\widetilde{p}_{ij}\widetilde{p}_{ji}\; D_{\widetilde{p}_{ij}}(f_{i},f_{j})+(\widetilde{p}_{ij}-\widetilde{p}_{ji})^2\\
   \\
   &=& 1-\diy\frac{4p_ip_j}{p_i+p_j}\int \frac{f_i(\bx)f_j(\bx)}{p_if_i(\bx)+p_jf_j(\bx)}\;\rd\bx\\
   \\
   &=& 1-\diy\frac{4}{p_i+p_j}\int \frac{p_if_i(\bx)p_jf_j(\bx)/(f^m(\bx))^2}{\big(p_if_i(\bx)+p_jf_j(\bx)\big)/f^m(\bx)}f^m(\bx)\;\rd\bx\\
   \\
   &=&1-\diy\frac{4}{p_i+p_j}\bbE_\BX\left[\frac{a_ia_j}{a_i+a_j}\right],
\end{array}\end{equation}
where $a_i=P(i|\bx)=p_if_i/f^{(m)}$. Therefore the RHS in (\ref{lowerbound.2.tight}) can be written as
\begin{equation}
    \diy\frac{1}{m}\diy\sum\limits_{i=1}^{m-1}\sum\limits_{j=i+1}^m (p_i+p_j)\left[\diy 1- \sqrt{1-\diy\frac{4}{p_i+p_j}\bbE_\BX\left[\frac{a_ia_j}{a_i+a_j}\right]}\right]. 
\end{equation}
Furthermore the LHS in (\ref{lowerbound.2.tight}) can be rewrite in terms of $a_i$ and $a_j$ as
\begin{equation}
\diy\frac{m-1}{m}\left[ 1 - \left(1 - 2\frac{m}{m-1} \bbE_\BX\left[\sum_{i=1}^{m-1} \sum_{j=i+1}^{m} a_i a_j\right]\right)^{1/2}\right]. 
\end{equation}
Note that since $\diy\sum\limits_{i}^m p_i=1$, we have $\diy\sum\limits_{i=1}^{m-1}\sum\limits_{j=i+1}^m (p_i+p_j)=m-1$, so that it is sufficient to show that 
\begin{equation}\label{eq2:Thm.tightLB2}\begin{array}{l}
    \diy\diy\sum\limits_{i=1}^{m-1}\sum\limits_{j=i+1}^m \diy \left((p_i+p_j)^2-\diy 4(p_i+p_j)\bbE_\BX\left[\frac{a_ia_j}{a_i+a_j}\right]\right)^{1/2}\\
    \\
    \qquad \geq \diy(m-1) \left(1 - 2\frac{m}{m-1} \bbE_\BX\left[\sum_{i=1}^{m-1} \sum_{j=i+1}^{m} a_i a_j\right]\right)^{1/2}. 
\end{array}\end{equation}
In addition we have \begin{equation}\label{eq.neq}\diy\sum\limits_{i=1}^{m-1}\sum\limits_{j=i+1}^m(...)^{1/2}\geq \left(\diy\sum\limits_{i=1}^{m-1}\sum\limits_{j=i+1}^m ...\right)^{1/2},\end{equation}
then from (\ref{eq2:Thm.tightLB2}), we need to prove that
\begin{equation}\label{eq3.Thm.tightLB2}\begin{array}{l}
    \diy\sum\limits_{i=1}^{m-1}\sum\limits_{j=i+1}^m \diy (p_i+p_j)^2-\diy 4\diy\sum\limits_{i=1}^{m-1}\sum\limits_{j=i+1}^m(p_i+p_j)\left[\frac{a_ia_j}{a_i+a_j}\right]\\
    \\
    \qquad\geq \diy(m-1)^2 - 2m(m-1) \sum_{i=1}^{m-1} \sum_{j=i+1}^{m} a_i a_j.\\
    \\
    \qquad=\diy(m-1) \diy\sum\limits_{i=1}^{m-1}\sum\limits_{j=i+1}^m \diy (p_i+p_j) - 2m(m-1) \sum_{i=1}^{m-1} \sum_{j=i+1}^{m} a_i a_j.
\end{array}\end{equation}
The following inequality implies (\ref{eq3.Thm.tightLB2})
\begin{equation}\label{eq.final.Thm9}
    (p_i+p_j)-4\left[\frac{a_ia_j}{a_i+a_j}\right]\geq m-1-\diy\frac{2m(m-1)}{p_i+p_j}\; a_ia_j.
\end{equation}
We know that $p_i+p_j\in(0,1)$ and $a_i+a_j\in(0,1)$ and since $\diy\sum_{l=1}^m p_l=1$ and $\diy\sum_{l=1}^m a_l=1$. One can check that for $m\geq 2$ the inequality (\ref{eq.final.Thm9}) holds true. This proves our initial claim in (\ref{lowerbound.2.tight}). 

\subsection{Proposition~\ref{prop:1}}
By assuming the equality of the bound, we can reduce to the following equality:
\begin{equation*}
  \sum_{i=1}^{m-1} \sum_{j=i+1}^m p_i p_j \int f_i(\bx)f_j(\bx)\left(\frac{1}{p_if_i(\bx) + p_jf_j(\bx)} - \frac{1}{f^m(\bx)} \right) \text{d} \bx = 0.
\end{equation*}
The above equality implies that either $m=2$ or for $m>2$ $(\bS^{i} \cup \bS^{(j)}) \cap (\cup_{k \neq
  i,j} \bS^{(k)})=\emptyset$.
Assume that $m > 2$. Note that the summands are always non-negative, due to the fact that $f^m(\bx)
\ge p_if_i(\bx) + p_jf_j(\bx)$. Therefore, in order for the equality to hold,
each summand must equal 0, that is:
\begin{equation*}
  \int f_i(\bx)f_j(\bx)\left(\frac{1}{p_if_i(\bx) + p_jf_j(\bx)} - \frac{1}{f^m(\bx)} \right) \text{d} \bx = 0.
\end{equation*}
Equivalently 
\begin{equation*}
  \int f_i(\bx)f_j(\bx)\left(\frac{\diy\sum\limits_{k\neq i,j}p_kf_k}{f^m(\bx)\left(p_if_i(\bx) + p_jf_j(\bx)\right)}  \right) \text{d} \bx = 0.
\end{equation*}
This can only occur when $f_i(\bx)$ or(and) $f_j(\bx)$ is 0 whenever $f_k > 0$ for
any $k \neq i,j$, or equivalently: $(\bS^{i} \cup \bS^{(j)}) \cap (\cup_{k \neq
  i,j} \bS^{(k)})=\emptyset$.

\subsection{Proposition~\ref{prop:2}}
By assuming the equality of the bound and following arguments in Subsection D, we reduce to the following equality:
\begin{equation}\label{eq2:Thm.tightLB3}\begin{array}{l}
    \diy\diy\sum\limits_{i=1}^{m-1}\sum\limits_{j=i+1}^m \diy \left((p_i+p_j)^2-\diy 4(p_i+p_j)\; \delta_{ij}\right)^{1/2}\\
    \\
  \qquad= \diy(m-1) \left(1 - 2\frac{m}{m-1} \sum_{i=1}^{m-1} \sum_{j=i+1}^{m} \delta^m_{ij}\right)^{1/2}. 
\end{array}\end{equation}
We know that equality occurs in (\ref{eq.neq}) iff $m=2$. 
Hence (\ref{eq2:Thm.tightLB3}) turns into 
\begin{equation}\label{eq2:Thm.tightLB4}\begin{array}{l}
    \diy\diy\sum\limits_{i=1}^{m-1}\sum\limits_{j=i+1}^m \diy (p_i+p_j)\left[(p_i+p_j)-\diy 4\; \delta_{ij}\right]\\
    \\
  \qquad= \diy \diy\sum\limits_{i=1}^{m-1}\sum\limits_{j=i+1}^m \diy (p_i+p_j)\left[(m-1)- \frac{2m(m-1)}{p_i+p_j} \delta^m_{ij}\right]. 
\end{array}\end{equation}
This implies that 
\begin{equation}\begin{array}{l}
    \diy\diy\sum\limits_{i=1}^{m-1}\sum\limits_{j=i+1}^m \diy (p_i+p_j)\Big[(p_i+p_j)-\diy 4\; \delta_{ij}\\
    \\
    \qquad\qquad-(m-1)+\diy\frac{2m(m-1)}{p_i+p_j} \;\delta^m_{ij}\Big]=0. 
\end{array}\end{equation} 
Note that the summands are always non-negative, due to the arguments in the proof of Theorem \ref{thm.lower.tight.2}. 
Therefore, assuming the equality of the bound reduces to 
\begin{equation}\label{prof.eq}
(p_i+p_j)-\diy 4\; \delta_{ij}-(m-1)+\diy\frac{2m(m-1)}{p_i+p_j} \;\delta^m_{ij}=0.\end{equation}
This equality holds iff $m=2$. The necessary condition is trivial to infer, we discuss the sufficient condition that is if \ref{prof.eq} holds then $m=2$. Assume $m\neq2$ or equivalently $m>2$ then we show that the LHS of (\ref{prof.eq}) is positive. The condition that the LHS of (\ref{prof.eq}) is greater than one is equivalent to: 
\begin{equation*}
\diy\frac{2m}{(p_i+p_j)} \;\delta^m_{ij}-\diy \frac{4}{m-1}\; \delta_{ij}> 1-\frac{(p_i+p_j)}{m-1}.\end{equation*}
We know that $0<\diy\frac{(p_i+p_j)}{m-1}<1$ therefore it is sufficient to show that for $m>2$,
\begin{equation*}
\diy\frac{2m}{(p_i+p_j)} \;\delta^m_{ij}-\diy \frac{4}{m-1}\; \delta_{ij}> 1.\end{equation*}
Furthermore since $(p_i+p_j)\in(0,1)$,  it is sufficient to show the following inequality: 
\begin{equation}\label{proof.equ2}
\diy 2m \;\delta^m_{ij}-\diy \frac{4}{m-1}\; \delta_{ij}> 1.\end{equation}
Now for small $m>2$ one may assume that $\delta^m_{ij}\eqsim \delta_{ij}$ and since $\diy 2m -\diy \frac{4}{m-1}> 1$ for all $m>2$ we then conclude the proof. For large $m$ such that $\delta^m_{ij}\ll\delta_{ij}$ and $2m \gg \diy\frac{4}{m-1}$ hence (\ref{proof.equ2}) holds true and the proof is completed. 

  

\def\bbE{\mathbb{E}}
\def\beqq{\begin{equation*}}
\def\eeqq{\end{equation*}}
\def\beq{\begin{equation}}
\def\eeq{\end{equation}}
\def\BY{{\mathbf{Y}}}
\def\by{{\mathbf{y}}}
\subsection{Theorem \ref{thm:5}}
Let $\mathbf{X}= \{(\bx_i, y_i)\}_{i=1}^n$ be an i.i.d. $m$-multiclass labeled sample. Let $N_{n_k}$ be Poisson variables with mean $n_k=\diy\sum_{i=1}^n I(y_i=k)$, for $k=1,\dots,m$ and independent of one another and of $\BX^{(k)}=\{(\bx_i,y_i)\}_{i=1,y_i=k}^n$. Now let $\overline{\BX}^{(k)}=\{(\bx_i,y_i)\}_{i=1,y_i=k}^{N_{n_k}}$, $k=1,\ldots,m$ be the Poisson process with FR statistic $\overline{\mathfrak{R}}^{(ij)}_{n_i,n_j}$ defined in Section~\ref{sec.estimate.delta} and constructed by global MST over $\bigcup\limits_{k=1}^m \overline{\BX}^{(k)}=\bigcup\limits_{k=1}^m \{(\bx_i,y_i)\}_{i=1,y_i=k}^{N_{n_k}}$. Following the arguments in \cite{HP} one yields that 
$$n^{-1}\bbE\left|\overline{\mathfrak{R}}^{(ij)}_{n_i,n_j}-\mathfrak{R}^{(ij)}_{n_i,n_j}\right|\rightarrow 0,$$
because of
\beqq \Big|\overline{\mathfrak{R}}^{(ij)}_{n_i,n_j}-\mathfrak{R}^{(ij)}_{n_i,n_j}\Big|\leq c_d \left(\diy\sum\limits_{k=1}^m \big|N_{n_k}-n_k\big|\right),\eeqq
where $c_d$ is the largest possible degree of any vertex in global MST over $\bigcup\limits_{k=1}^m \BX^{(k)}$, $\BX^{(k)}=\{(\bx_i,y_i)\}_{i=1,y_i=k}^n$. Hence it remains to prove that
\beq\label{thm5.proof.eq0}
\diy \frac{\bbE\left[\overline{\mathfrak{R}}^{(ij)}_{n_i,n_j}\right]}{2n}\rightarrow \delta_{ij}^m.
\eeq
\def\BZ{\mathbf{Z}}\def\bz{\mathbf{z}}
For $n_1^m:=(n_1,\dots,n_m)$ let $\BZ_1^{n_1^m},\BZ_2^{n_1^m},\dots$ be independent vectors with common densities $g^{(m)}_{n}(\bx)=\diy\sum\limits_{k=1}^m n_k f_k(\bx)/n$. Next let $K_n$ be an independent Poisson variable with mean $n$. Consider $\BZ_n=\left\{\BZ_1^{n_1^m}, \BZ_2^{n_1^m},\dots,\BZ_{K_n}^{n_1^m}\right\}$ a nonhomogeneous Poisson process of rate $\diy\sum\limits_{k=1}^m n_k f_k(\bx)$. Assign a mark from the set $\{1,2,\dots,m\}$ to each point of $\BZ_n$. A point at $\bx$, independently of other points, being assigned the mark $s$ with probability $n_{s} f_s\big/(\diy\sum\limits_{k=1}^m n_k f_k(\bx))$, for $t=1,\dots,m$. Let $\widetilde{\BX}^{(s)}_{n_s}$ denotes the set of points in $\BZ_n$ with mark $s$ for $s=1,2,\dots,m$ i.e. $\widetilde{\BX}^{(s)}_{n_s}=\{(\BZ_i,y_i)\}_{i=1,y_i=s}^{n_s}$. Introduce $\widetilde{\mathfrak{R}}^{(ij)}_{n_i,n_j}$ as the FR statistics for data set $\widetilde{\BX}^{(1)}_{n_1}\cup \widetilde{\BX}^{(2)}_{n_2}\cup\dots \cup \widetilde{\BX}^{(m)}_{n_m}$, applying the global MST and counting edges connecting a point with mark $i$ to a point with mark $j$. Using the marking theorem $\widetilde{\BX}^{(s)}_{n_s}$, for all $s=1,\dots,m$ are independent Poisson process with the same distribution as $\overline{\BX}^{(s)}$. Therefore we prove (\ref{thm5.proof.eq0}) for $\widetilde{\mathfrak{R}}^{(ij)}_{n_i,n_j}$, see \cite{HP}, once again. Given points of $\BZ_n$ at $\bx$ and $\bz$, the probability that they have marks $i$ and $j$
\beqq 
W^{(m)}_{n_i,n_j}(\bx,\bz):=\diy\frac{n_{i} f_i(\bx) n_{j} f_j(\bz)+n_{j} f_j(\bx) n_{i} f_i(\bz)}{\left(\diy\sum\limits_{k=1}^m n_k f_k(\bx)\right)\left(\diy\sum\limits_{k=1}^m n_k f_k(\bz)\right)}.
\eeqq
Then for $1\leq i<j\leq m$
\begin{equation}\label{eq2:1.1}\begin{array}{l}
\bbE\left[\widetilde{\mathfrak{R}}^{(ij)}_{n_i,n_j}|\BZ_{n}\right]\\
\quad=\diy\mathop{\sum\sum}_{\ 1\leq t<l\leq K_{n}} W^{(m)}_{n_i,n_j}(\BZ^{n_1^m}_t,\BZ^{n_1^m}_l)
\times \mathbf{1}\big\{(\BZ^{n_1^m}_t, \BZ^{n_1^m}_l)\in \mathfrak{F}(\BZ_{n})\big\},
\ena\end{equation}
here $\mathfrak{F}(\BZ_{n})$ represents the global MST over nodes in $\BZ_{n}$. 
Hence, we have
\beqq\begin{array}{l}
\bbE\left[\widetilde{\mathfrak{R}}^{(ij)}_{n_i,n_j}|\BZ_{n}, \right]\\
\quad=\diy\mathop{\sum\sum}_{\ 1\leq t<l\leq K_{n}} W^{(m)}_{n_i,n_j}(\BZ^{n_1^m}_t,\BZ^{n_1^m}_l) \mathbf{1}\big\{(\BZ^{n_1^m}_t, \BZ^{n_1^m}_l)\in \mathfrak{F}(\BZ_{n})\big\}. 
\ena\eeqq
Further, set
\beqq
W^{(m)}(\bx,\bz):=\diy\frac{p_ip_j\big(f_i(\bx)f_j(\bz)+f_j(\bx)f_i(\bz)\big)}{\left(\diy\sum\limits_{k=1}^m p_k f_k(\bx)\right)\left(\diy\sum\limits_{k=1}^m p_k f_k(\bz)\right)}.
\eeqq
One can check that $W^{(m)}_{n_i,n_j} \rightarrow W^{(m)}$ and  they range in $[0,1]$. Next by taking expectation from (\ref{eq2:1.1}), we can write 
\begin{equation}\begin{array}{l}
\diy\bbE\left[\widetilde{\mathfrak{R}}^{(ij)}_{n_i,n_j}\right]\\
=\diy\bbE\mathop{\sum\sum}_{\ 1\leq t<l\leq K_{n}} W^{(m)}(\BZ^{n_1^m}_t,\BZ^{n_1^m}_l) \mathbf{1}\big\{(\BZ^{n_1^m}_t, \BZ^{n_1^m}_l)\in \mathfrak{F}(\BZ_{n})\big\}\\
\qquad\qquad +o(n).\ena\end{equation} 
By taking into account the non-Poisson process $\BZ'_{n}=\left\{{\BZ}^{n_1^m}_1,{\BZ}^{n_1^m}_2, \dots, {\BZ}^{n_1^m}_{{n}}\right\}$ and the fact that $\bbE\big[\big|\diy\sum\limits_{k=1}^m N_{n_k}-n\big|\big]=o(n)$, one yields:
\begin{equation} \begin{array}{l}
\diy \bbE\left[\widetilde{\mathfrak{R}}^{(ij)}_{n_i,n_j}\right]\\
\qquad=\diy\bbE\mathop{\sum\sum}_{\ 1\leq t<l\leq n} W^{(m)}(\BZ^{n_1^m}_t,\BZ^{n_1^m}_l) \mathbf{1}\big\{(\BZ^{n_1^m}_t, \BZ^{n_1^m}_l)\in \mathfrak{F}(\BZ'_{n})\big\}\\
\qquad\qquad\qquad+o(n).\ena\end{equation}
Also, we can write that $g^{(m)}_{n}(\bx)\rightarrow g^{(m)}(\bx)$ where $g^{(m)}(\bx)=\diy\sum\limits_{k=1}^m p_k f_k(\bx)$. Consequently by proposition 1 in \cite{HP}, we have
\beqq \begin{array}{l}
\diy \frac{\bbE\left[\widetilde{\mathfrak{R}}^{(ij)}_{n_i,n_j}\right]}{n}\rightarrow\int W^{(m)}(\bx,\bx) g^{(m)}(\bx)\rd\bx\\
\\
\qquad \qquad \qquad =2\diy\int \diy\frac{p_{i} p_{j} f_i(\bx) f_j(\bx)}{\diy\sum\limits_{k=1}^m p_k f_k(\bx)}\rd\bx. 
\ena\eeqq
This completes the proof.

\label{sec:refs}
\bibliographystyle{IEEEbib}
\bibliography{refs}
\end{document}